\documentclass[12pt]{iopart}

\usepackage{iopams}  
\usepackage{graphicx}
\usepackage{braket}
\usepackage{comment}
\usepackage{xcolor}

\begin{document}

\title[Control, comp. and coex. of eff. mag. orders by int.  in BEC with high-Q cav.]{Control, competition and coexistence of effective magnetic orders by interactions in  Bose-Einstein condensates with high-Q cavities}

\author{Brahyam R\'\i os-S\'anchez and Santiago F. Caballero-Ben\'\i tez}

\address{Departamento de F\'\i sica Cu\'antica y Fot\'onica, LSCSC-LANMAC, Instituto de F\'{i}sica, Universidad Nacional Aut\'{o}noma de M\'{e}xico, Ciudad Universitaria, Mexico City, C. P. 04510, Mexico}
\ead{scaballero@fisica.unam.mx}
\vspace{10pt}

\begin{abstract}
Ultracold atomic systems confined in optical cavities have been demonstrated as a laboratory for the control of quantum matter properties and analog quantum simulation. Often neglected, but soon amenable to manipulation in a new generation of experiments, we show that atomic many-body interactions allow additional control in the cavity driven self-organization of effective spinor Bose-Einstein condensates (BEC). We theoretically show that a rich landscape of magnetic ordering configurations emerges.  This can be  controlled by modifying the geometry of the light-fields in the system with the interplay of two-body interactions and the cavity induced interactions. This leads to competition scenarios and phase separated dynamics. Our results show that it is possible to tailor on demand configurations possibly useful for analog quantum simulation of magnetic materials with highly controllable parameters in a single experimentally realistic setup.
\end{abstract}

%
\vspace{2pc}

\section{Introduction}
\begin{figure}[t!]
    \centering    \includegraphics[scale = 0.23]{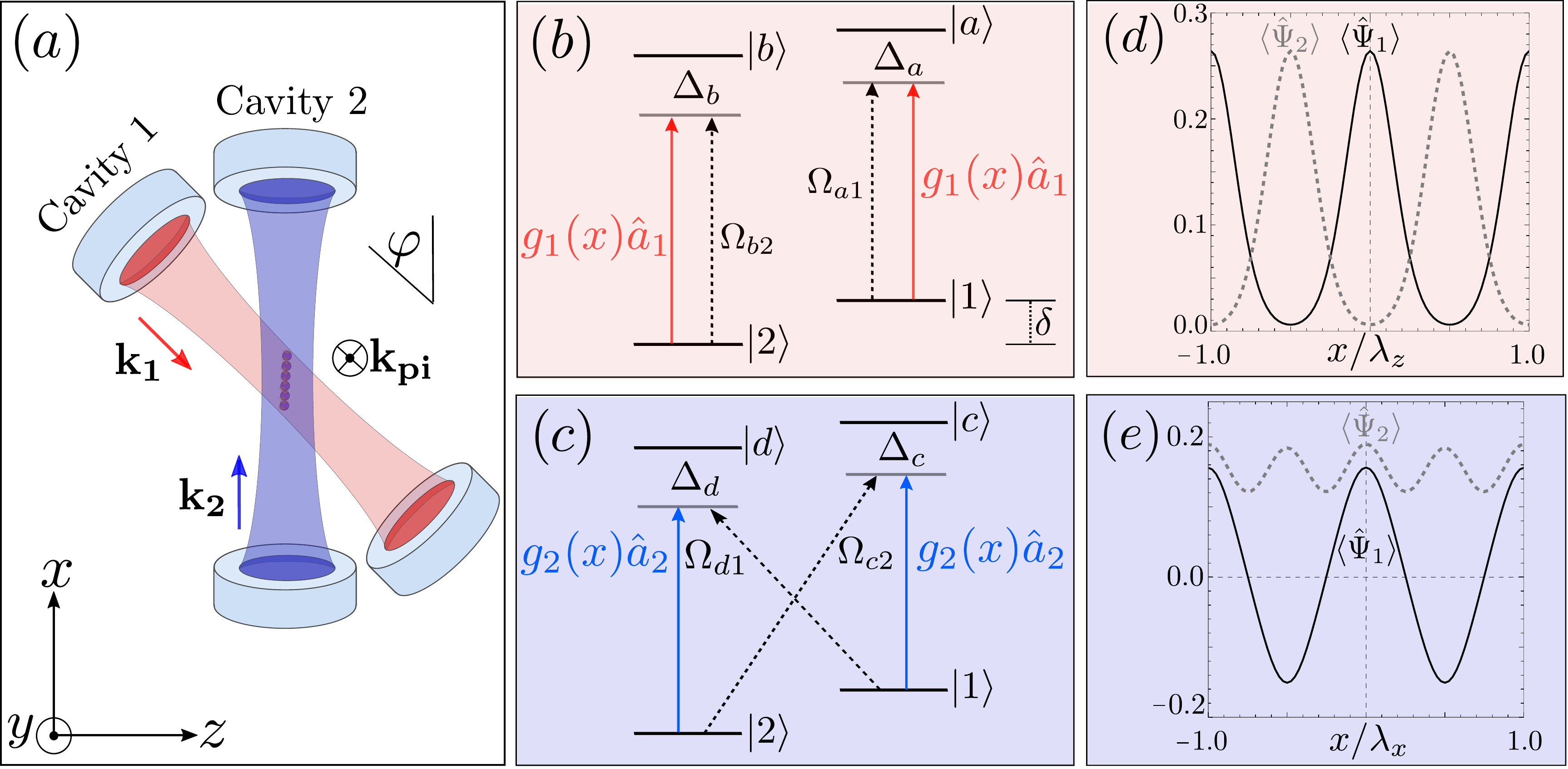}
    \caption{(a) Schematic representation of a 1D BEC trapped in the optical potentials generated by two-crossed cavities. Each cavity mediates the coupling between the ground state $\ket{2}$, the excited state $\ket{1}$ and a set of two different auxiliary levels via (b) two direct dipolar transitions with Rabi frequency  $g_{1}(x) =g_{01}\cos{(\mathbf{k_1} \cdot \mathbf{r})}$ and (c) two double-$\Lambda$ Raman transitions with $g_{2}(x) =g_{02}\cos{(\mathbf{k_2} \cdot \mathbf{r})}$. The dotted lines correspond to transitions with constant frequencies $\Omega_{\mu i}$ induced by transversal laser-pumping from the $y$ direction.  {Figures (d) and (e) show the typical  spatial redistribution of atoms in the states $\ket{1}$ and $\ket{2}$ induced by the processes described in (b) and (c), respectively. Here $\lambda_x = 2\pi k^{-1}_x$ and $\lambda_z = 2\pi k^{-1}_z$, being $k_x\equiv|\mathbf{k_2}|$ and $k_z \equiv |\mathbf{k_1}|\cos{(\varphi)}$. The pump laser and cavity frequencies, $\omega_{pi}$ and $\omega_{ci}$ respectively, are far red-detuned from all atomic transitions.} 
    }
    \label{fig_systemScheme}
\end{figure}

The interplay between the physics of quantum optics and ultracold atoms have proven to be a fertile ground for both theoretical and experimental research. The experimental versatility of the optical control techniques enables the engineering of ultracold systems as quantum simulators \cite{quantumSimulator, quantumSimulatorPRX, Cirac2012}, where the physics of condensed matter can be widely explored. From a theoretical perspective, the interaction between radiation and ultracold matter can give rise to novel phases due to the emergence of dynamic optical potentials and effective long-range interactions between atoms, which can alter the critical properties of macroscopic matter. 
\\
In particular, the dynamics of the multi-component Bose Einstein condensates coupled to quantum and classical states of light display a wide range of intriguing phenomena, including the emergence of magnetic \cite{Caballero2022, magneticMoire, magneticToolbox, spinTextures, spinDensityOrder}, spatial \cite{cristallineDipolar, Tricriticality, stripeSupersolid} and temporal orders \cite{ timeCrystal1, timeCrystal2}. A recurring characteristic in the phenomenology of the mentioned systems is the role played by short-range two-body interactions in order to ensure the stability of the emerging phases. Despite this, the effects of short-range interactions have been seldom explored in the physics of cavity-assisted self-organization in effective spinor BEC \cite{LightInducedQuasicrystal, cavitySpinOrbit, spinOrbit, phaseSeparationSelforganization, competingOrders, chiralTopo}. 
\\
In this work, we show that the short-range interactions influence the competition and coexistence of self-organization orders in an elongated BEC trapped at the intersection of two high-finesse optical cavities. This leads to a rich landscape that can be exploited to do analog quantum simulation of magnetic systems in a single setup Fig. \ref{fig_systemScheme}. By means of properly tuning the atomic collisions and cavity Rabi frequencies it is possible to engineer any possible scenario of magnetic ordering on demand, see Fig. \ref{fig_coex_table} and  section \ref{sec_separation}. Moreover, it would be easy to measure the emergence of magnetic order due to the relationship between the light in each cavity and the corresponding self-organized atomic state.
\\
The model under consideration is built on recent experimental observation of competing order parameters \cite{Morales2018, crossedCavities2017} and cavity-induced BEC self-organization \cite{PRLSelforganization2017}, by incorporating the short-range interactions in order to study their effect on the criticality of the self-organization phase transition with the possibility of magnetic domain formation exploiting the emergence of phase segregation of the atomic components. 
\\
This paper is presented as follows: In section \ref{sec_model}, we introduce the model used to describe the self-organization competition of the coherently-coupled two component interacting BEC. We briefly describe the structure of the steady state configurations by analyzing the associated semi-classical energy functional. In section \ref{sec_stability}, we discuss the dynamical stability of the homogeneous steady state and provide an expression for the excitation spectrum. Subsequently, we identify the impact of short-range interactions on the self-organization thresholds. In section \ref{sec_competition}, we report our numerical findings concerning to the coexistence and competition of self-organization orders as function of the short-range interaction couplings and the ratio of the cavity mode wavelengths. In section \ref{sec_separation}, we discuss the effects of the density segregation on the self-organization transition and the conformation of local magnetic domains. Finally, in section \ref{sec_conclusions}, we conclude our study and mention some directions for further research.

\section{Model}\label{sec_model}
Consider a collection of bosonic atoms confined in the total optical potential generated by a setting of two crossed optical cavities, similar to the setting in~\cite{Morales2018,crossedCavities2017,Higgs}. The transversal frequencies of an external harmonic trap restrict the spatial dynamics of the atoms in an quasi-1D configuration along the $x$ axis, as shown in Fig. \ref{fig_systemScheme} (a). The atomic internal degrees of freedom consist in a set of six energy levels, with the ground state subspace $\ket{1}$ and $\ket{2}$ being involved in both cavity-assisted transitions. The direct transition $\ket{1}\rightarrow \ket{a}$ ($\ket{2}\rightarrow \ket{b}$) is mediated by a combination of {two counter-propagating and transversely pumped laser fields of frequencies $\omega_{p1}$, $\omega_{p2}$ and the field of the Cavity 1 in a standing wave configuration}. The associated Rabi couplings are $\Omega_{a1}$ ($\Omega_{b2}$) and $g_{1}(x) =g_{01}\cos{(\mathbf{k_1} \cdot \mathbf{r})}$, respectively, as shown in Fig. \ref{fig_systemScheme} (b). Note that in this scheme the ground states are effectively coupled via the cavity mode. On the other hand, the two-photon processes $\ket{1} \leftrightarrow \ket{2}$ are the result of a double-$\Lambda$ transitions produced by {two auxiliary laser fields $\omega_{p3}$, $\omega_{p4}$ } and the Cavity 2 mode with Rabi couplings $\Omega_{d1}$, $\Omega_{c2}$ and $g_{2}(x) =g_{02}\cos{(\mathbf{k_2} \cdot \mathbf{r})}$. This level scheme is illustrated in Fig. \ref{fig_systemScheme} (c). The double-$\Lambda$ transitions depicted here can be implemented by using two hyperfine states (for instance, $\ket{2} = \ket{F=1, m_F =1}$, $\ket{1} = \ket{F=1, m_F = 0}$) which are split by an external magnetic field orthogonal to the Cavity 2 axis. By aligning the polarization vector of the pumping laser fields parallel to the Cavity 2 axis, a superposition of circular polarization components is produced, enabling selective excitation of atomic transitions with $\Delta m_F = \pm 1$ via cavity-assisted Raman processes, as demonstrated in \cite{Zhiqiang:17, Barret:17}.  Other possibilities, beyond the standing wave configuration, i.e. traveling or running waves, are possible too \cite{HelmutRev, HelmutRing, Caballero2015}, which lead to complex mode functions. {We consider all the pumped laser fields propagating orthogonal to the BEC extension, then the spatial variation of these fields is not relevant to the dynamics of the confined atoms.} Moreover, as the atoms are tightly confined on the $x$ direction, only the projection of the cavity wavevectors, named $k_z \equiv |\mathbf{k_1}|\cos{(\varphi)}$ and $k_x\equiv|\mathbf{k_2}|$ influence the spatial redistribution of atoms, being $0 < \varphi < \pi/2$ the angle between the cavity axes. Note that the quantity $\xi_{zx} = k_z / k_x$ characterizes the emergent lattice spacing of the effective optical potential. {However, the pump laser frequencies are chosen such that the two-photon Raman transitions depicted in Fig. \ref{fig_systemScheme} (c) are close to resonance, meaning $\omega_{c2} - \omega_{p3} \approx \omega_{p4} - \omega_{c2} \approx \delta$ \cite{PRLSelforganization2017}. Similarly, for the direct transitions in \ref{fig_systemScheme} (b) we consider $\omega_{p1} \approx \omega_{p2} \approx \omega_{c1}$.} Here, we use the subscripts $x$ and $z$ for the wavevectors related to the  spin projections of  emergent self-organized  magnetic orders in the emergent  quantization axis, as discussed in the following sections.
{Each of the light-matter interaction schemes presented in Fig. \ref{fig_systemScheme} (b) and Fig. \ref{fig_systemScheme} (c)  drives independently a distinct spatial redistribution of the atomic density, as illustrated in in Fig. \ref{fig_systemScheme} (c) and Fig. \ref{fig_systemScheme} (d), respectively. The interplay and competition between these emergent spatial structures, in conjunction with two-body atomic interactions, underpins the emergence of a rich variety of effective magnetic orders, as discussed below.}
\\
We employ the second quantization formalism to describe the dynamics of the presented atom-field system. Let $\hat{\Psi}_\mu(x, t)\equiv \hat{\Psi}_\mu$ the annihilation operator of one atom in the $\ket{\mu}$ at the position $x$ for $\mu = 1,2,a,...,d$. Similarly, denote by $\hat{a}_i$ the annihilation operator of a photon in the cavity mode $\cos{(k_i x)}$ ($i = 1,2$). These field operators satisfy the usual bosonic commutation relations, $[\hat{\Psi}_\mu(x,t), \hat{\Psi}^\dagger_\nu(x^\prime, t)]=\delta_{\mu,\nu}\delta(x-x^\prime)$ and $[\hat{a}_{i}, \hat{a}^{\dagger}_{j}] = \delta_{i, j}$. 
Under the rotating-wave approximation, the Hamiltonian governing the dynamics reads $\hat{H} = \hat{H}^\mathrm{A} + \hat{H}^\mathrm{C} + \hat{H}^\mathrm{AC}_1 + \hat{H}^\mathrm{AC}_2$, where:
\begin{equation}
\hspace{-1.0 cm}\hat{H}^\mathrm{A} = \sum_\mu\int \rmd x\hat{\Psi}_\mu^\dagger\bigg(-\frac{\hbar^2}{2m}\partial^2_x - \hbar\Delta_\mu \bigg)\hat{\Psi}_\mu
   	+\frac{1}{2}\sum_{ij=1,2}U_{ij}\int \rmd x \hat{\Psi}_i^\dagger\hat{\Psi}_j^\dagger\hat{\Psi}^{\phantom{\dagger}}_j\hat{\Psi}^{\phantom{\dagger}}_i, 
\end{equation}
describes the atomic configuration decoupled of the radiation fields. It consists of the kinetic energy of the atoms of mass $m$, the atomic frequency detuning $\Delta_\mu$ and the short-range two-body interactions with strength $U_{ij}$, which are related to scattering length $a_{ij}$ via $U_{ij} = 2\pi \hbar \omega_\bot a_{ij}$, with $\omega_\bot$ being the transversal harmonic frequency confining the BEC. This strength corresponds to  the intra-species (for $i =j$) and inter-species (for $i\neq j$) collisions between atoms  in the ground state subspace. This is a reasonable approximation as the excited states have a low occupation probability. The cavity photons are described by: 
\begin{equation}
    \hat{H}^\mathrm{C} = -\hbar\Delta_{c1} \hat{a}^\dagger_{1} \hat{a}^{\phantom{\dagger}}_{1} - \hbar\Delta_{c2} \hat{a}^\dagger_{2} \hat{a}^{\phantom{\dagger}}_{2},
\end{equation}
where $\Delta_ {ci}$ stands for the cavity detunings. It is worth to mention that all $\Delta_\mu$ and $\Delta_{ci}$ are defined as their correspondent bare frequencies relative to those of the pump laser fields (see \ref{ap:modelDetails} for details). In this work we consider the regime of red detuning $\Delta_\mu <0$ and $\Delta_{ci} <0$. The processes depicted in Fig. \ref{fig_systemScheme} (b) are described by the Hamiltonian:
\begin{equation}
    \hspace{-1.0 cm}\hat{H}^\mathrm{AC}_1 = \hbar\int \rmd x  \bigg[\big(g_1(x)\hat{a}_1 + \Omega_{b2}\big)\hat{\Psi}^\dagger_b\hat{\Psi}^{\phantom{\dagger}}_2 +  \big(g_1(x)\hat{a}_1 + \Omega_{a1}\big)\hat{\Psi}^\dagger_a\hat{\Psi}^{\phantom{\dagger}}_1 + \mathrm{H.c.} \bigg],
\end{equation}
while those in Fig. \ref{fig_systemScheme} (c) are considered in:
\begin{equation}
	\hspace{-1.0 cm}\hat{H}^\mathrm{AC}_2  = \hbar\int \rmd x \bigg[ g_2(x)\hat{a}_2\big(\hat{\Psi}^\dagger_d\hat{\Psi}^{\phantom{\dagger}}_2 + \hat{\Psi}^\dagger_c\hat{\Psi}^{\phantom{\dagger}}_1\big) + \Omega_{c2}\hat{\Psi}^\dagger_c\hat{\Psi}^{\phantom{\dagger}}_2 + \Omega_{d1}\hat{\Psi}^\dagger_d\hat{\Psi}^{\phantom{\dagger}}_1 +\mathrm{H.c.} \bigg].
\end{equation}
\noindent
The full Hamiltonian encloses rich physical scenarios emerging from the interplay between the complex internal atomic structure and the short- and global-range interactions. Indeed, the two body interactions $U_{ij}$ serve as a mechanism to ensure the stability of superfluid phases in ultracold atoms. In the case of multi-component BEC, it can provide scenarios for mutually-trapping and multi stability \cite{selfbound, Dalafi_2013}. On the other hand, the cavity optical potentials can give rise to crystalline density distributions from which the driven-dissipative supersolid-like phases are predicted to occur \cite{Deng-2023, Qin-2022, DrivenSupersolid}. The potential experimental realizations are in the direction of the control techniques developed in \cite{Morales2018, crossedCavities2017, Higgs}.

\subsection{Effective two-component model}\label{subsec_effectiveModel}
\begin{figure}[t!]
    \centering    \includegraphics[scale = 0.275]{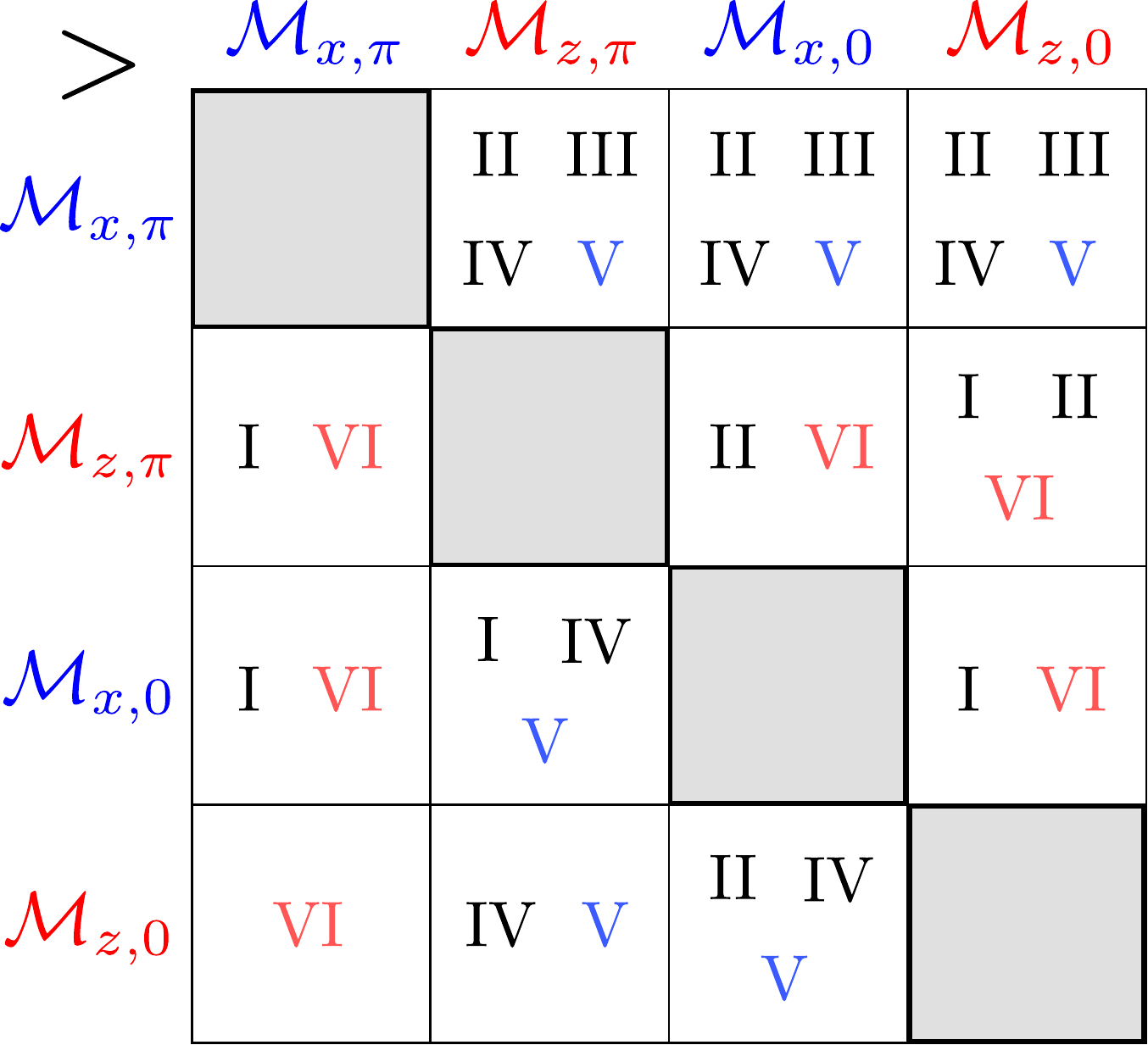}
    \caption{Coexistence of magnetic orders. The table presents the comparison between pairs of magnetic orders, where those on the columns are larger than the ones on the rows. The roman numeral indicates the steady-states wavefunctions depicted in Fig. \ref{fig_wavefunctions1-3} and Fig. \ref{fig_wavefunctions4-6}. The table illustrates that any combination is possible by carefully tuning the short range effective interactions and the cavity induced interactions.}
    \label{fig_coex_table}
\end{figure}

In the ground state at $T=0$ at the level of meanfield,  we introduce the order parameters that describe the condensation in the ground state subspace as $\langle \hat{\Psi}_i \rangle = \sqrt{N}\psi_{i}(x, t) \equiv \sqrt{N}\psi_i$ ($i = 1, 2$), with $N$ the number of atoms. {  This formalism assumes that the number of atoms per lattice site is large, making statistical averaging of interactions a reliable approximation.} The adiabatic elimination of the excited subspace and the cavity fields in equations give rise to the following set of effective coupled Gross-Pitaevskii equations for the condensate wavefunctions (see \ref{ap:modelDetails} for details):

\begin{equation}
   \eqalign{
   \hspace{-2cm}
   \rmi\hbar\partial_{t} \psi_1 = \bigg(-\frac{\hbar^2}{2m}\partial^2_{x} + \frac{\hbar\delta}{2}+N U^{\phantom{*}}_{11}
   |\psi^{\phantom{*}}_1|^2
   +N U^{\phantom{*}}_{12 }
   |\psi^{\phantom{*}}_2|^2
   &- N \hbar J_z \mathcal{M}_{z, \pi} \cos{(k_{z}x)}\bigg)\psi^{\phantom{*}}_{1}\\
 &- N \hbar J_x \mathcal{M}_{x, \pi}\cos{(k_{x}x)}\psi_{2},
 }
    \label{eq_GPE1}
\end{equation}

\begin{equation}
     \eqalign{
     \hspace{-2cm} 
    \rmi\hbar\partial_{t} \psi_2 = \bigg(-\frac{\hbar^2}{2m}\partial^2_{x} - \frac{\hbar\delta}{2}+N U^{\phantom{*}}_{22}
    |\psi^{\phantom{*}}_2|^2+N U^{\phantom{*}}_{12}
    |\psi^{\phantom{*}}_1|^2 &+ N \hbar J_z \mathcal{M}_{z, \pi}\cos{(k_z x)}\bigg)\psi^{\phantom{*}}_{2}\\
    &- N \hbar J_x \mathcal{M}_{x, \pi}\cos{(k_x x)}\psi_{1}.
    }
    \label{eq_GPE2}
\end{equation}
where $\hbar\delta$ is the Stark-shifted energy detuning between the atoms in the ground subspace, and $J_{\sigma}>0$ are the effective two-photon Rabi frequency mediating the coupling of the atoms with the photons of the cavity mode with spatial profile $\cos{(k_{\sigma} x)}$ ($\sigma = x, z$), see Appendix B for details. For simplicity, a symmetrical inter-species interaction $U_{21} = U_{12}$ has been assumed. {Since the number of atoms $N$ explicitly appears in the coupled the Gross-Pitaevskii equations (\ref{eq_GPE1}) and (\ref{eq_GPE2}), the condensate wavefunctions are normalized such that $\int dx (n_1 + n_2) = 1$, where $N n_i = N\psi^*_{i}\psi^{\phantom{*}}_i$ is the density of the condensed atoms in the $i=1,2$ component \cite{Rogel-Salazar_2013}}. By employing the Schwinger representation of angular momentum algebra, the optical potential is generated when the  pseudo-spin polarization per atom $s_x(x) \equiv \langle \hat{S}_x \rangle /N = \psi^{*}_1\psi^{\phantom{*}}_2 + \psi^{*}_2\psi^{\phantom{*}}_1$, or  the density polarization per atom $s_z (x) \equiv \langle \hat{S}_z \rangle /N  = n_1 - n_2$ are spatially modulated. Thus, $s_z$ defines an emergent quantization axis of our analog magnetic system, in an abstract sense. For this reason, it is convenient to define the 
order parameters,
\begin{equation}
    \mathcal{M}_{\sigma, q}=\int \rmd x\cos{(q x/l_\sigma)}s_\sigma(x), \qquad l_\sigma = \lambda_\sigma/2.
    \label{eq_orderParameters}
\end{equation}
Whenever $\mathcal{M}_{\sigma, \pi}\neq 0$, the total density of the BEC obeys the Bragg condition, which enhances the dispersion of photons into the cavity. This process results in an effective positive feedback mechanism in which the { steady-state} super-radiant dispersion of photons is given by,
{
\begin{equation}
	\langle\hat{a}^{ss}_1\rangle \equiv \alpha_z = \frac{\tilde{g}_1 N}{\tilde\Delta_1+i\kappa_1}\mathcal{M}_{z, \pi}, \qquad \langle\hat{a}^{ss}_2\rangle\equiv \alpha_x = \frac{\tilde{g}_2 N}{\tilde\Delta_2+i\kappa_2}\mathcal{M}_{x, \pi}
\end{equation}
with $\tilde{g}_2=g_{02}\Omega_{c2}/\Delta_c$ and  $\tilde{g}_1={g_{01}\Omega_{a1}}/{\Delta_a}$ the effective light matter couplings, $\kappa_i$ the decay rate  for each cavity mode and
\begin{equation}
	\tilde{\Delta}_{1/2}=\Delta_{c1/c2} - \int \rmd x g^2_{1/2}(x)\bigg(\frac{\langle \hat{\Psi}^\dagger_1\hat{\Psi}^{\phantom{\dagger}}_1\rangle}{\Delta_{c/a}} + \frac{\langle \hat{\Psi}^\dagger_2\hat{\Psi}^{\phantom{\dagger}}_2 \rangle}{\Delta_{d/b}} \bigg)
\end{equation}
the effective cavity detuning; see Appendix B for details. 
}Therefore, there is dynamically generation of density modulations in the condensate which induces back action to the photon dispersion and  vice versa. This leads to spatial self-organization \cite{selforganization2010}, but now with magnetic ordering. The system effectively realizes ferromagnetic  (FM) or anti-ferromagnetic configurations (AFM). The order parameters in equation (\ref{eq_orderParameters}) are reminiscent of the staggered $\mathcal{M}_{\sigma, \pi}$ (AFM) and direct $\mathcal{M}_{\sigma, 0}$ (FM) magnetization per site on lattice models \cite{Caballero2022}. As a consequence, the light buildup in each of the cavities gives the  direct information of the magnetic order present in the system. This can be readily measured by placing photo detectors at each arm and the particular spatially dependent behavior of each atomic component could be isolated by time of flight imaging, extracting the spatial frequencies involved to reconstruct $s_\sigma(x)$. It is possible by tuning the system parameters to accommodate combinations of coexistence of orders, where at least two $\mathcal{M}_{\sigma, q}$ are nonzero, as shown in Fig. \ref{fig_coex_table}. In the scenarios where $\mathcal{M}_{\sigma, \pi}\neq0$ the condensate wavefunctions realizes AFM orders with periodic spatial modulations of extended periodicity when $k_z /k_x$ is a rational number. 

\subsection{Semiclassical energy functional}

\begin{figure}[t!]
	\centering    \includegraphics[scale = 0.2]{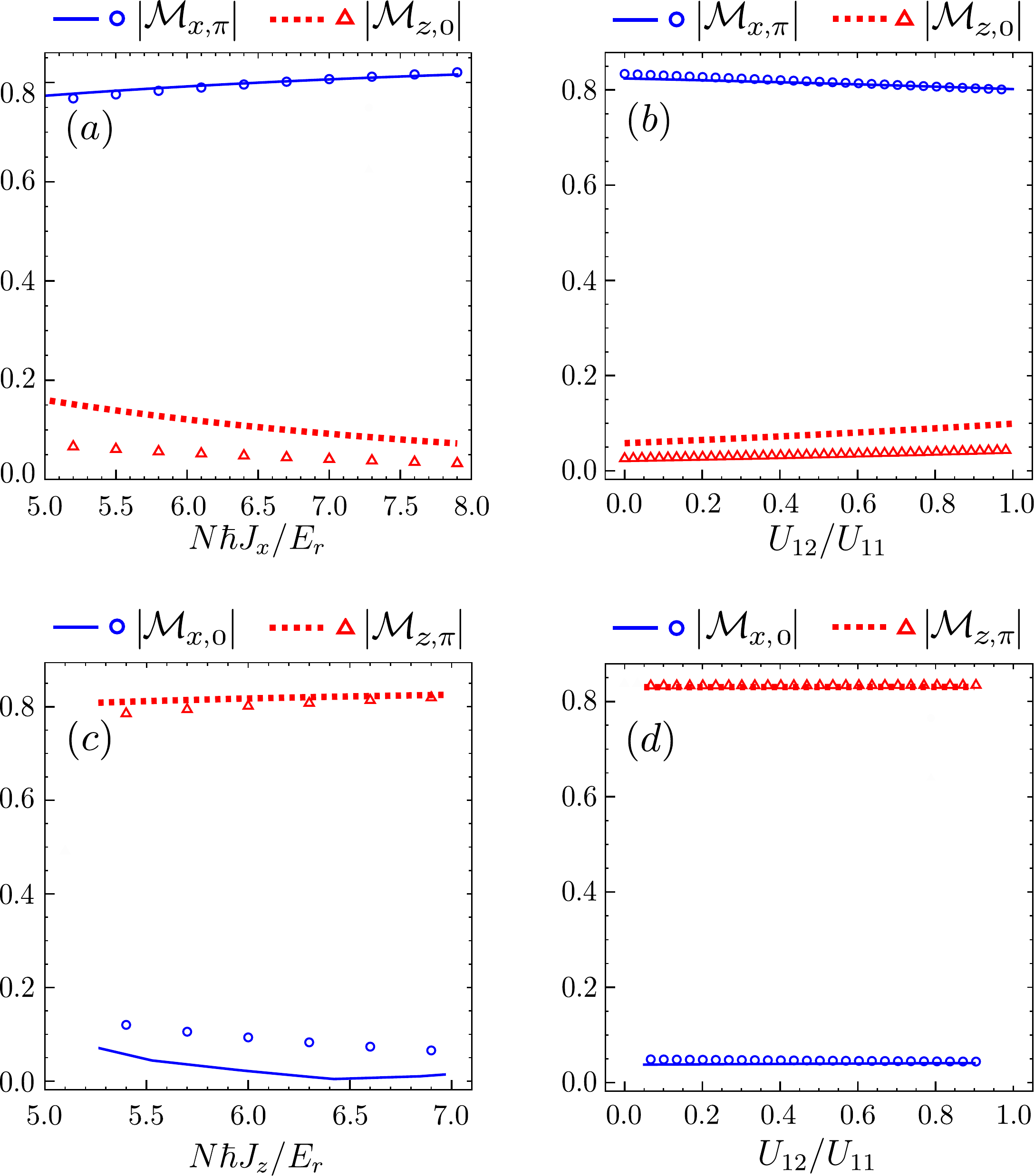}
	\caption{{
			Magnetic orders under the two mode approximation. The lines correspond to the variational calculations using the ansatz in equations (\ref{eq_twoModeMx}) and (\ref{eq_twoModeMz}), while markers to the numerical solution of equations (\ref{eq_GPE1}) and (\ref{eq_GPE2}). 
			The saturation of $\mathcal{M}_{x, \pi}$ and the persistence of the $\mathcal{M}_{z, 0}$ is shown for (a) $U_{12}/U_{11} = 0.4$ and (b) $N\hbar J_x = 8 E_r$, where $\hbar\delta = 0.5 E_r$. Similarly, the saturation of $\mathcal{M}_{x, \pi}$ is depicted for (c) $U_{12}/U_{11} = 0.4$ and (d) $N\hbar J_x = 8 E_r$, with $\hbar\delta = 0.5 E_r$ and $k_z/k_x = 3/4$. The rest of the parameters are $NU_{11} = NU_{22} = 1.75 E_r L_x$ and $L_x = 25.6 \lambda_x$. The energy and length scales used corresponds to the recoil energy $E_r = \hbar^2 k_x^2 /2m$ and wavelength $\lambda_x = 2\pi k^{-1}_x$, respectively.}}
	\label{fig_StimationMagnetization}
\end{figure}

Due to the coherent coupling and elastic scattering between condensate atoms, the total energy and the number of atoms in the condensate are conserved. The former can be expressed by accounting for the internal energy of the atoms, the two body contact interactions and the long-range interactions mediated by the optical potentials:
\begin{equation}
  \eqalign{
  	\hspace{-1.5cm}
    E =\frac{\hbar^2}{2m}\int \rmd x \big(|\partial_x \psi_1|^2  + |\partial_x \psi_2|^2\big) 
    &+N\int \rmd x \bigg(\frac{U_{11}}{2}n^2_{1}+\frac{U_{22}}{2}n^2_{2}+U_{12} n_1 n_2\bigg)\\
    &+\frac{\hbar\delta}{2}\mathcal{M}_{z, 0}-\frac{N\hbar J_x}{2}\mathcal{M}^2_{x, \pi} - \frac{N\hbar J_z}{2}\mathcal{M}^2_{z, \pi}.
    }
    \label{eq_energyFunctional}
\end{equation}
The analysis of each term in the energy functional reveals the competition between different spatial orders in the system: In absence of dynamically generated optical potentials ($\mathcal{M}_{\sigma, \pi} = 0$), the energy functional possess $U(1)\times U(1)$ symmetry due to the global phase invariance of $\psi_1$ and $\psi_2$. Additionally, the system supports configurations with continuous translation symmetry $\mathcal{T}$ if $U_{12} < \sqrt{U_{11}U_{22}}$ or density-segregated configurations if $U_{12} > \sqrt{U_{11}U_{22}}$ and $\delta = 0$ \cite{pitaevskiiBook, twoComponentArxiv}. On the other hand, when the two body interactions are negligible in comparison with the optical potential energy scale, our model effectively reduces to the one studied in \cite{PRLSelforganization2017} if $J_z = 0$. In this situation, the system spontaneously breaks the $U(1)\times U(1)$ and $\mathcal{T}$ symmetries as $J_x$ exceeds the self-organization threshold. As consequence, { a momentum transfer of $p_x = \hbar k_x$ is imparted to the atoms in component $\ket{1}$. From the analysis of the magnetization order parameter in equation (\ref{eq_orderParameters}), with $\sigma = x$, follows that the self-organization transition restricts the the momentum transfer to atoms in component $\ket{2}$ to be even multiples of $\hbar k_x$, as only even powers of the cavity mode $\cos{(k_x  x)}$ have non-zero spatial averages. As a result, the BEC wavefunctions $\psi_1(x)$ and $\psi_2(x)$ exhibit spatial modulations of wavelength $\lambda_x$ and $\lambda_x/2$ respectively, as depicted in Fig. \ref{fig_systemScheme} (e)}. Further, the relative phase between the BEC wavefunctions is energetically locked such  that $\cos{\big(\arg(\psi_1) - \arg(\psi_2) \big)} = \pm 1.$ Finally, if $J_x = 0$, the system undergoes a self-organization transition preserving the $U(1)\times U(1)$ symmetry { resulting from the collective momentum transfer of $p_x = \hbar k_z$ to each BEC component, as shown in Fig. \ref{fig_systemScheme} (d).}
\\
In order to gain physical insight, we explore the $\mathcal{M}_{\sigma,\pi}$ AFM orders under the two mode approximation. Deep in the $\mathcal{M}_{x,\pi}$ AFM phase, {the steady state consistent with the provided description of the BEC wavefunctions in the self-organization regime is given by:}
\begin{equation}
	\hspace{-2cm}
    \psi_1 = \sqrt{\frac{2}{L_x}}
    c_1\cos{(k_x x)}, \hspace{0.5cm} \psi_2 = \sqrt{\frac{1}{L_x}}\big(c_0 + \sqrt{2}c_2\cos{(2 k_x x)}\big), \hspace{0.5cm} c_0^2 + c_1^2 + c_2^2 = 1,
    \label{eq_twoModeMx}
\end{equation}
with $L_x$ being the BEC length. The minimization of the energy functional in equation  (\ref{eq_energyFunctional}) shows that this AFM phase strongly suppresses the $\mathcal{M}_{x,0}$ and $\mathcal{M}_{z,\pi}$ orders due to both scale as $(k_xL_x)^{-1}$ (for details see \ref{sec_appendix_twoMode}). Interestingly, the FM order $\mathcal{M}_{z,0}$ persist, as shown in Fig. \ref{fig_StimationMagnetization} (a) and (b). A similar behavior is observed in the deep $\mathcal{M}_{z,\pi}$ order, where the minimization of the energy functional given the ansatz 
\begin{equation}
	\hspace{-2.4cm}
    \psi_1 = \sqrt{\frac{1}{L_x}}\big(a_0 + \sqrt{2}a_1\cos{(k_z x)}\big), \hspace{0.35cm}\psi_2 = \sqrt{\frac{1}{L_x}}\big(b_0 + \sqrt{2}b_1\cos{(k_z x)}\big),\hspace{0.3cm}\sum_i(a^2_i + b^2_i)=1.
    \label{eq_twoModeMz}
\end{equation}
The above predicts the suppression of the $\mathcal{M}_{x, \pi}$ order, as shown in Fig. \ref{fig_StimationMagnetization} (c) and (d).
These competing and coexistence of magnetic orders are confirmed by the numerical simulation of the coupled GPEs, as discussed below.

\section{Stability of the homogeneous ground state} \label{sec_stability}

\begin{figure}[t!]
	\centering    
	\includegraphics[scale = 0.25]{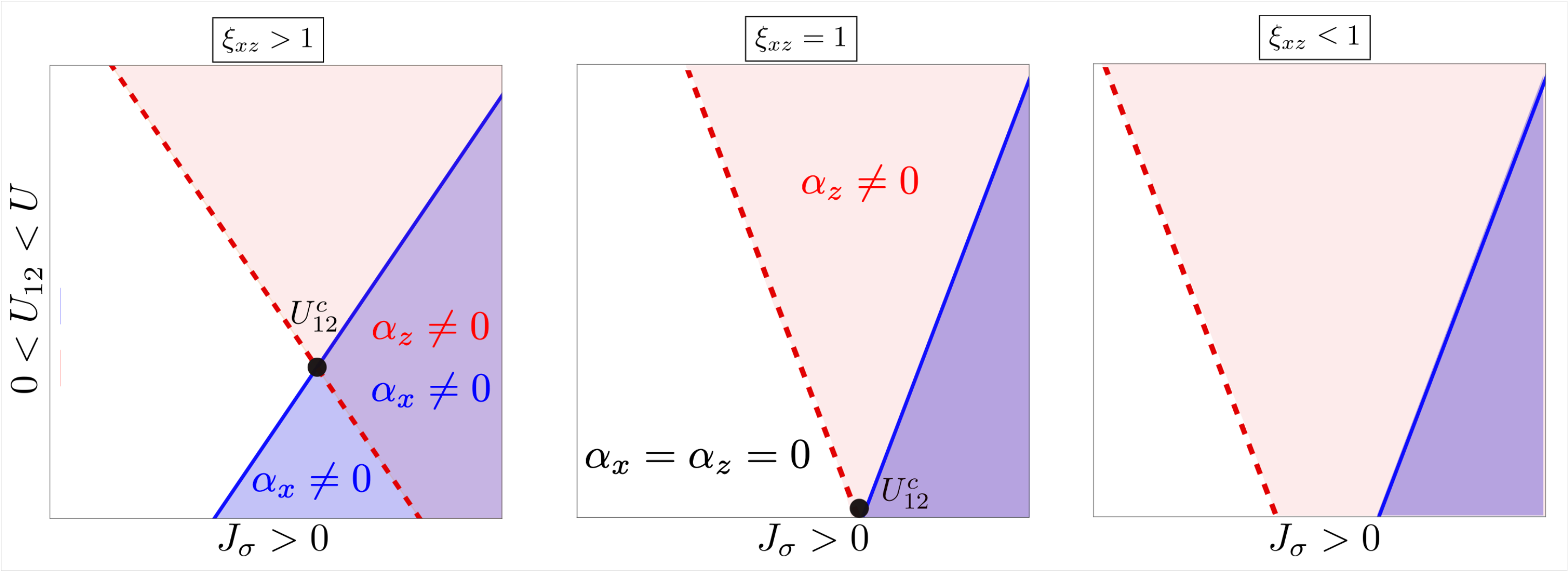}
	\caption{
		Phase diagram of the independent super-radiant transitions. The light amplitude in each cavity of the system is $\alpha_\sigma$ with $\sigma\in\{x,z\}$; the diagram of the system is in Fig. \ref{fig_systemScheme}. The blue-solid and red-dotted lines correspond to the self-organization thresholds $J^c_x$ and $J^c_z$, respectively. The coexistence and competition of self-organization orders occur in the region where both field amplitudes are nonzero. The dot refers to the value of $U_{12}$ from which the long-wavelength mode requires more energy to trigger the super-radiant transition compared to the short-wavelength mode.}
	\label{fig_stabilityDiagram}
\end{figure}

In order to understand the order competition in the model, we study the excitation spectrum by considering density fluctuations in the ground state configuration given by $\psi_i(x, t) = (\phi_{i}(x)+\delta\psi_i(x,t))\rme^{-\rmi\mu_{i} t /\hbar}$, where $\mu_{i}$ is the chemical potential for each BEC component. At first order in density fluctuations $\delta\psi_i$, the equations (\ref{eq_GPE1}) and (\ref{eq_GPE2}) result in the linear form $\rmi\hbar\partial_t \Lambda(x,t) = \mathcal{L}\Lambda(x,t) $, with $\Lambda(x,t) = (\delta\psi^{\phantom{*}}_1, \delta\psi^*_1, \delta\psi^{\phantom{*}}_2, \delta\psi^*_2)^T$. The linear stability analysis was shown to yield the same expression for the excitation spectrum as derived from Bogoliubov theory for weakly interacting Bose-Einstein condensates \cite{BogoliubovCoh2003}. In the miscible regime $\sqrt{U_{11}U_{22}}>U_{12}$, the ground state below the self-organization threshold is defined by homogeneous density distributions $\phi_i(x)=\sqrt{n_i}$ with the associated chemical potentials $\mu_1= \hbar\delta/2 + NU_{11}n_1 + NU_{12}n_2$ and $\mu_2 = -\hbar\delta/2 + NU_{22}n_2 + NU_{12}n_1$. Note that $\mu_1 = \mu_2$ due to chemical equilibrium in the ground state. Following a similar procedure as the presented in \cite{excitations}, the stability matrix $\mathcal{L}$ can be calculated analytically in the momentum space by the following anzats,
\begin{equation}
    \delta\psi_{i}(x,t) = \sum_k \cos{(kx)}\bigg[\rme^{-\rmi\omega(k)t}\alpha_{i, k} + \rme^{\rmi\omega^*(k)}\beta^*_{i, k} \bigg],\label{eq_densityFluctuations}
\end{equation}
where $\hbar\omega(k)$ are the eigenvalues of $\mathcal{L}$. The spatial dependence in the preceding ansatz is considered to include the information of the spatial profile of the cavity modes. In addition, we allow $\omega(k)$ to be a complex number in order to include unstable configurations emerging in the system. This follows as the stability matrix is in general non-Hermitian. Taking into account that the spatial order parameters in equation (\ref{eq_orderParameters}) are zero in the homogeneous ground state, the stability matrix reads:
\begin{equation}
   \mathcal{L} = \left( \begin{array}{cccc}
    E_k + V_1 & V_1 & V_{12} & V_{12} \\
    -V_1 & -(E_k + V_1) & -V_{12} & -V_{12} \\
    V_{12} & V_{12} & E_k + V_2 & V_2 \\
    -V_{12} & -V_{12} & -V_2 & -(E_k + V_2) 
    \end{array} \right)
    \label{eq_stabilityMatrix}
\end{equation}
Here, $E_k = \hbar^2 k^2 / 2m$ corresponds to the energy spectrum of the non-interacting BEC,
\begin{equation}
    V_i/N = U_{i i}n_i - \delta_{k, \pm k_z}\frac{\hbar J_z}{2n}n_i - \delta_{k, \pm k_x}\frac{\hbar J_x}{2n}n_{3 - i}, 
\end{equation}
\begin{equation}
    V_{12}/N = \sqrt{n_1 n_2}\bigg(U_{12} - \delta_{k, \pm k_x}\frac{\hbar J_x}{2n} + \delta_{k, \pm k_z}\frac{\hbar J_z}{2n}\bigg)
\end{equation}
and $n = n_1 + n_2$ is the BEC density per atom. In the balanced configuration defined by $U_{11} = U_{22} = U$ and by the energy degeneration of the BEC components, $\hbar\delta =0$, the ground state is characterized by $n_1 = n_2 = n/2$. This yields a compact expression for the excitation spectrum above the homogeneous ground state given by the eigenvalues of $\mathcal{L}$ in equation (\ref{eq_stabilityMatrix}):
\begin{equation}
    \hbar^2\omega^2_{x}(k) = E_k\bigg[E_k + N(U + U_{12})n - N\hbar J_{x}\delta_{k, \pm k_x}\bigg],
    \label{eq_densityMode}
\end{equation}
\begin{equation}
    \hbar^2\omega^2_{z}(k) = E_k\bigg[E_k + N(U-U_{12})n - N\hbar J_{z}\delta_{k, \pm k_z}\bigg].
    \label{eq_spinMode}
\end{equation}
For $J_\sigma = 0$, the two branches of the excitation spectrum reduce to the well-known spin ($\hbar\omega_z$) and density ($\hbar\omega_x$) modes of a two component BEC coupled by contact interactions \cite{pitaevskiiBook}. In previous works \cite{abad2013, reviewCoupledBECS} was shown that coherently coupled BECs exhibit an energy gap at $k=0$ in the spin mode signaling a spontaneous density polarization for a critical value of the coupling constant. In our study case, the spatial-dependent coherent coupling does not open a gap at $k=0$ in the excitation spectrum, not even in the limit $k_\sigma\rightarrow 0$. Nevertheless, in this long-wavelength limit the critical values of $U_{i j}$ at which the density segregation takes place are modified. Another interesting characteristic of the excitation spectrum in equations (\ref{eq_densityMode}) and (\ref{eq_spinMode}) is the fact that each of its branches is independently influenced by the Fourier transform of only one cavity mode. These allow the identification of two self-organization thresholds determined by the condition for $\omega_z(k_z)$ and $\omega_x(k_x)$ to acquire imaginary parts indicating the onset of exponential growth of density fluctuations in equation (\ref{eq_densityFluctuations}):
\begin{equation}
    N\hbar J^c_{x} = E_r + N(U + U_{12})n, 
    \label{eq_Jx_critic}
\end{equation}
\begin{equation}
   N\hbar J^c_{z} =  \xi^2_{zx}E^{\phantom{2}}_r + N(U - U_{12})n,
   \label{eq_Jz_critic}
\end{equation}
where $E_r =\hbar^2 k^2_x /2m$ is the recoil energy imparted on the atoms due to dispersion of photons in the cavity mode $\cos{(k_x x)}$ and $\xi_{zx} = k_z / k_x$. The difference in the critical values define stability zones of the homogeneous ground state depending on $\xi_{zx}$, $J_\sigma$ and $U_{12}$. For instance, there is a value of $U_{12}$ defined as
\begin{equation}
    U^c_{12} = \frac{1}{2Nn}(\xi^2_{zx}-1)E_r,
\end{equation}
from which the self-organization threshold in equation (\ref{eq_Jx_critic}) requires the highest activation energy despite being the longest wavelength mode. These results show that the destabilization of both excitation modes may induce density fluctuations leading to the competition of two self-organized configurations. 
\\
By experimentally manipulating accessible parameters as laser intensities ($J_\sigma$) and external magnetic fields ($U_{12}$), the competition between self-organization orders can be observed by measurement of the intra-cavity photon number or by resolving the BEC momentum distribution. This presents a promising avenue to study the stationary and dynamical effects of order competition over a wide range of parameters due to the extent of the stability zones in Fig. \ref{fig_stabilityDiagram}.

\begin{figure}[t!]
    \centering
   \includegraphics[scale = 0.2]{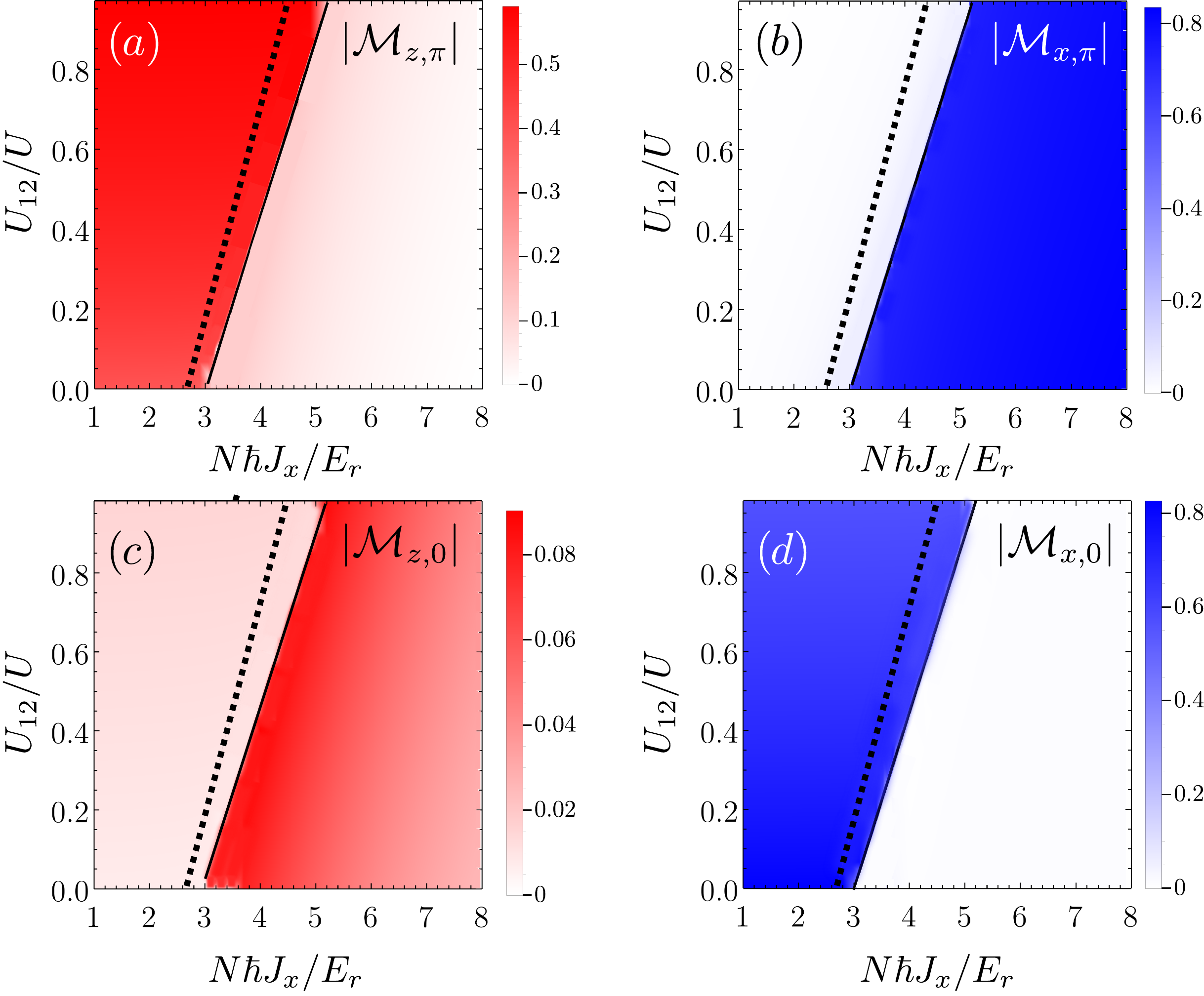}
         \caption{ 
    Staggered and direct magnetization as function of the Rabi frequency $J_x$ and $U_{12}/U$ for a fixed  value of $J_z > J^c_z$. The continuous line indicates the self-organization threshold for $\mathcal{M}_{x,\pi}$. The coexistence of magnetic orders is visualized in the region where both AFM orders $\mathcal{M}_{\sigma, \pi}$ and the FM $\mathcal{M}_{z,0}$ differ from zero, as shown in panels $(a,c)$ and in Fig. \ref{fig_slices}. On addition, the panels $(b,d)$ exhibit the competition of $\mathcal{M}_{\sigma,x}$ orders as mutual exclusion. The values of the remaining physical parameters are $N\hbar J_z = 5.5 E_r$, $\hbar \delta = 0 E_r$, $NUn = 1.75 E_r$ and $\xi_{zx} = 7/4$.The  dotted lines correspond to the analytical self-organization threshold for $\xi_{zx}=1$.
    } 
    \label{fig_orderParametersSevenFour}
\end{figure}

\section{Competition and coexistence of self-organization orders.} \label{sec_competition}
The steady state of the coupled Gross-Pitaevskii equations (\ref{eq_GPE1}) and (\ref{eq_GPE2}) was computed by implementing the imaginary-time evolution via split-step Fourier method \cite{split-step-1, split-step-2}. Owing to the imaginary-time evolution operator is real, the obtained wavefunctions resulting from the propagation of real-valued wavefunctions are also real-valued. Due to the two-photon Raman transitions, the arguments of the macroscopic wavefunctions  are energetically locked, such that $\cos{(\arg(\psi_1) - \arg(\psi_2))}=\pm 1$.

\begin{figure}[t!]
    \centering
     \includegraphics[scale = 0.2]{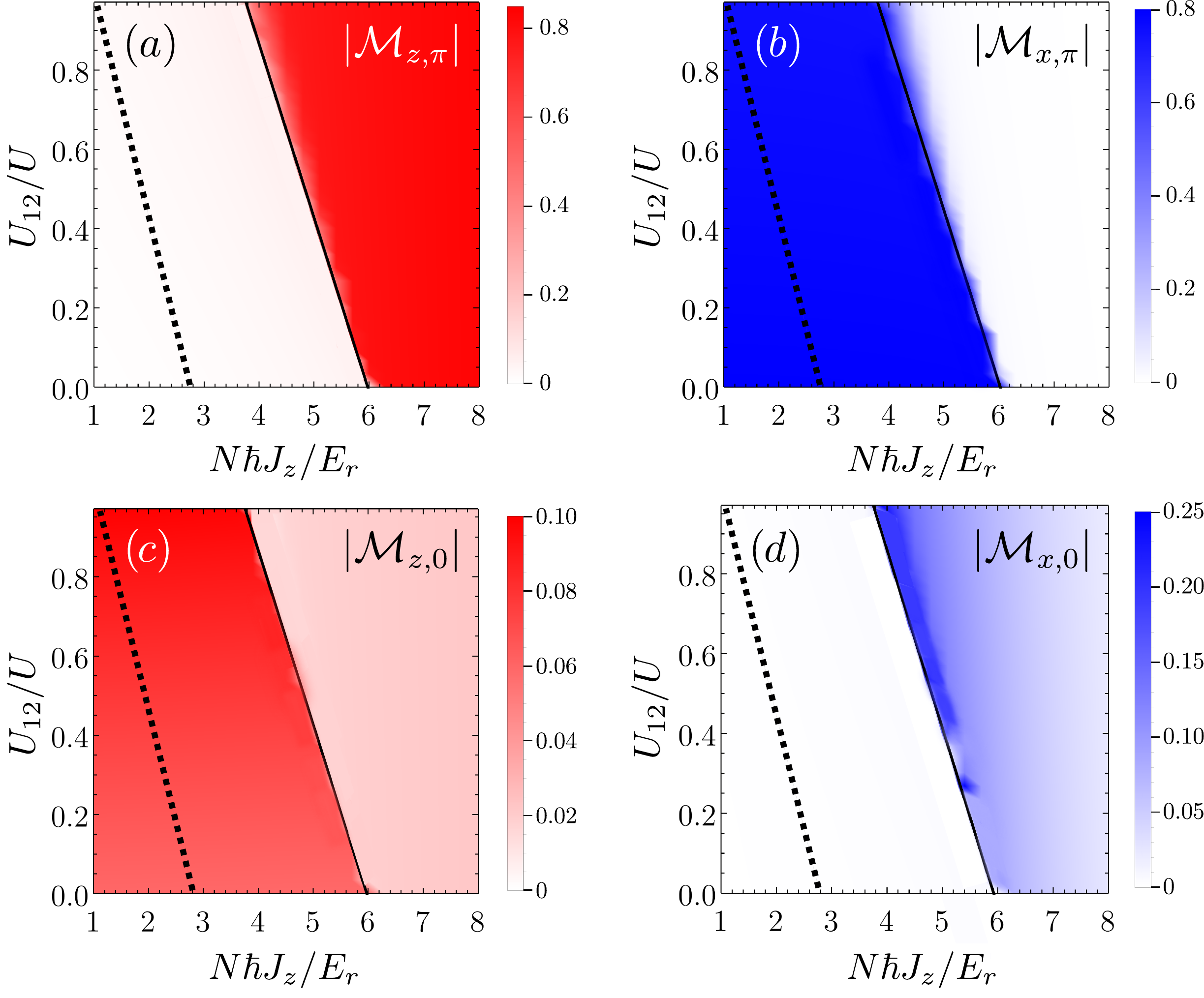}
         \caption{Staggered and direct magnetization as function of the Rabi frequency $J_z$ and $U_{12}/U$ for a fixed  value of $J_x > J^c_x$. The continuous line indicates the self-organization threshold for $\mathcal{M}_{z,\pi}$.
         The order competition is observed as the suppression of one of the AFM orders in panels $(a)$ and $(b)$. On contrast, the FM orders $\mathcal{M}_{\sigma,0}$ coexist with the AFM order $\mathcal{M}_{z,\pi}$, as shown in panels $c)$ and $d)$. The values of the remaining physical parameters are $N\hbar J_x = 5.5 E_r$, $\hbar \delta = 0 E_r$ and $\xi_{zx} = 3/4$. The  dotted lines correspond to the analytical self-organization threshold for $\xi_{zx}=1$.
    }
    \label{fig_orderParametersThreeFour}
\end{figure}
\noindent
From these results, the order competition is observed in the behavior of the order parameters defined in equation (\ref{eq_orderParameters}) as the values of the quantities $J_\sigma$ and $U_{12}$ are varied through the stability zones shown in Fig. \ref{fig_stabilityDiagram}. Concretely, Fig. \ref{fig_orderParametersSevenFour} depicts the steady-state magnetization as $J_x$ increases from $J_x < J^c_x$. The value of $J_z$ is fixed above its self-organization threshold. Initially, the system configuration is governed by the $J_z$-induced self-organization due to the $\mathcal{M}_{z, \pi}$ order parameter is nonzero. As consequence, the spatial modulations of the BEC wavefunctions are $\lambda_z$-periodic as expected from the optical potential $\cos{(k_z x)}$ (see Fig. \ref{fig_wavefunctions1-3} (a)). Once the self-organization threshold in equation (\ref{eq_Jx_critic}) is exceeded, the order competition is triggered and an abrupt change in both order parameters $\mathcal{M}_{x, \pi}$ and $\mathcal{M}_{z, \pi}$ occur \cite{orderParameters}. As $J_x$ increases, the order parameter $\mathcal{M}_{x, \pi}$ approaches to the saturation value while $\mathcal{M}_{z, \pi}$ tends to zero. Interestingly, this behavior is also observed in the direct magnetization order parameters, being the FM order $\mathcal{M}_{x,0}$ suppressed as $J_x$ increases. { The dotted lines in Figs. \ref{fig_orderParametersSevenFour} and \ref{fig_orderParametersThreeFour} correspond to the analytical self-organization threshold for $\xi_{zx}=1$. The coexistence and competence of the magnetic effective orders induce an energy shift in the self-organization threshold. }
\\
In the case $\xi_{zx}\leq 1$, the self-organization thresholds become more separated as $U_{12}\rightarrow U$, affecting the self-organization competition. This is the scenario shown in Fig. \ref{fig_orderParametersThreeFour}, where the roles of $J_x$ and $J_z$ are interchanged for clarity and completeness purposes. The key difference here is that the self-organization competition is not triggered as soon as $J_z$ exceeds the threshold in equation (\ref{eq_Jz_critic}). This may be attributed to the fact that the value of $J^c_z$ is lower enough compared with the value of $J_x > J^c_x$ to be energetically suppressed. Therefore, the density fluctuations arising from the self-organization transition are strongly diminished. It is only when $J_z = 3.4 J^c_z$ that the order competition takes place with a rapid saturation of $\mathcal{M}_{z, \pi}$ and a significant suppression of $\mathcal{M}_{x, \pi}$ occurring much faster compared to the case shown in Fig. \ref{fig_orderParametersSevenFour} (a).
\\

\begin{figure}[ht!]
    \centering
    \includegraphics[scale = 0.21]{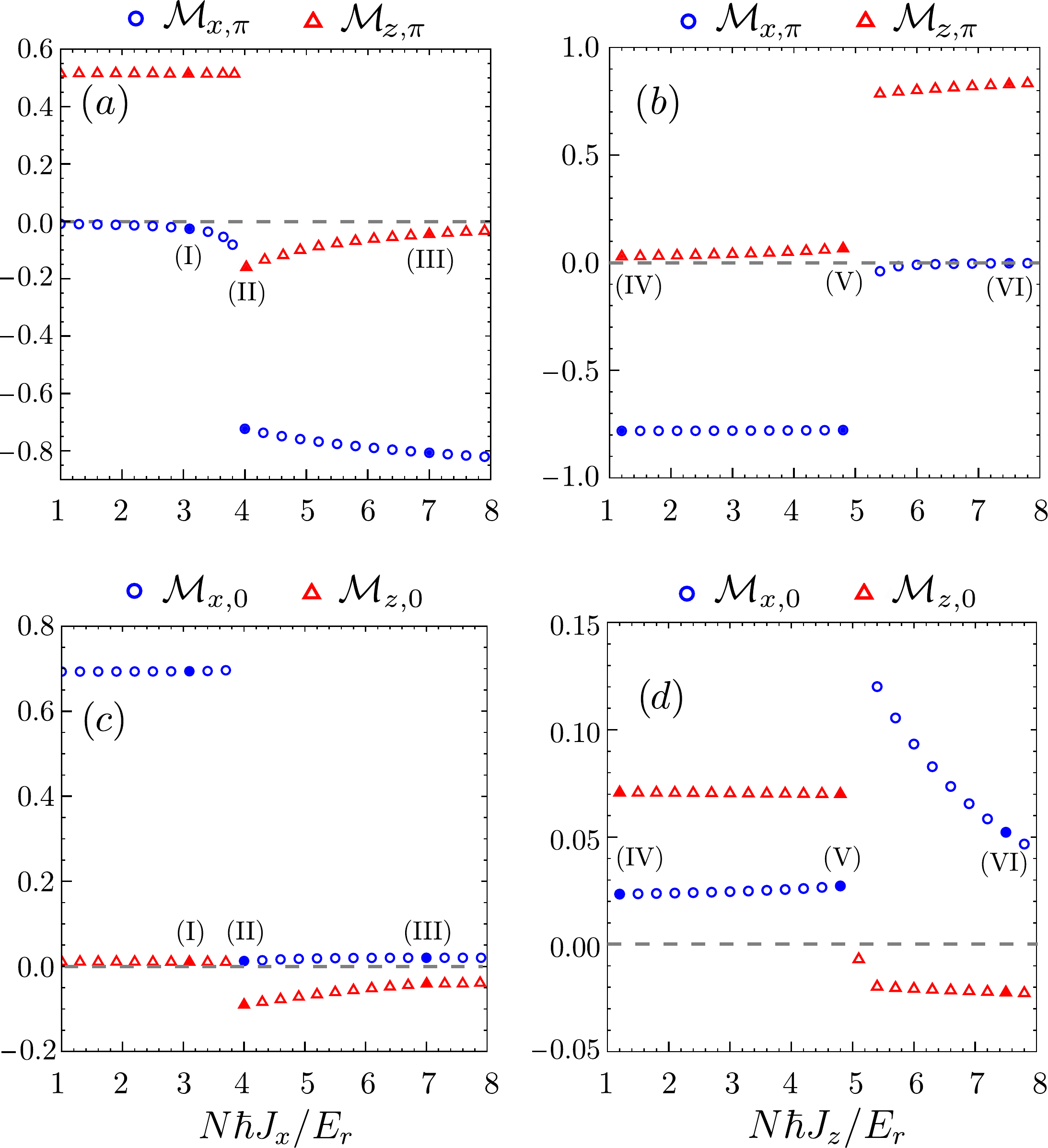}
    \caption{Slices of the magnetization phase diagrams shown in Fig. \ref{fig_orderParametersSevenFour} and Fig. \ref{fig_orderParametersThreeFour} for $U_{12}/U=0.4$. In panel $(a)$ the coexistence of AFM orders occurs in the region close to $N\hbar J_x > 4E_r$, while in $(b)$ is observed the suppression of one of the AFM orders. This behavior is also observed in the FM orders shown in $(c)$ and $(d)$. The Roman numerals indicate the BEC wavefunctions and their correspondent spin components depicted in Fig, \ref{fig_wavefunctions1-3} and Fig. \ref{fig_wavefunctions4-6}.
    }
    \label{fig_slices}
\end{figure}
\begin{figure}[ht!]
    \centering
    \includegraphics[scale = 0.3]{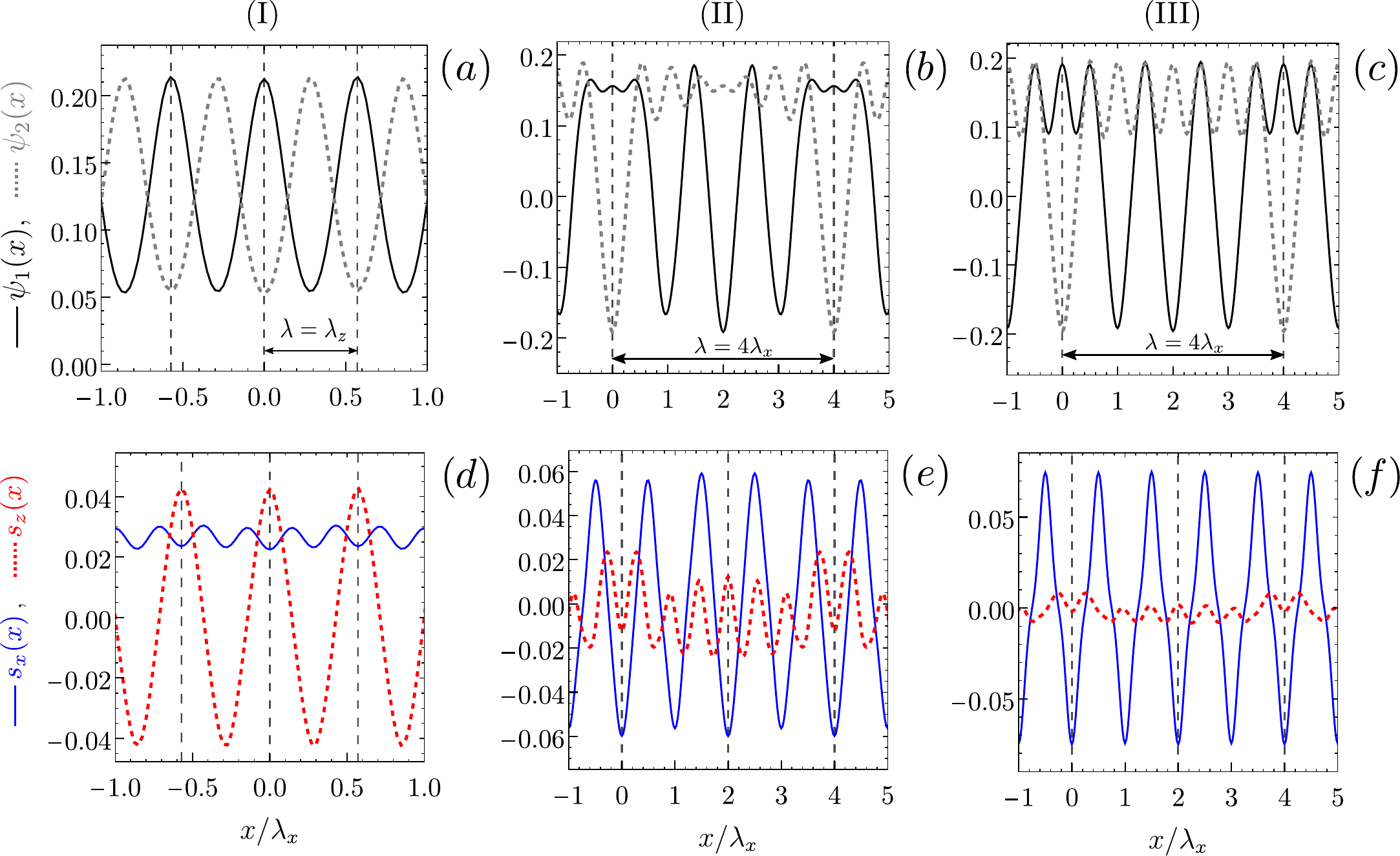}
    \caption{
    Steady-state wavefunctions and pseudo-spin components corresponding to the magnetic configurations indicated in Fig. \ref{fig_slices}. As $J_x$ increases, the initially-dominant AFM order $\mathcal{M}_{z,\pi}$ in $(a)$ and $(d)$ gives place to a configuration of extended periodicity $\lambda = 4\lambda_x$ in the region of AFM order coexistence, as shown in $(b)$ and $(c)$. Nevertheless, the magnetic order on the coexistence region becomes dominated by the antiferromagnetic behavior of $s_x(x)$, as shown in $(e)$ and $(f)$.
    
   }
    \label{fig_wavefunctions1-3}
\end{figure}
\begin{figure}[ht!]
    \centering
    \includegraphics[scale = 0.3]{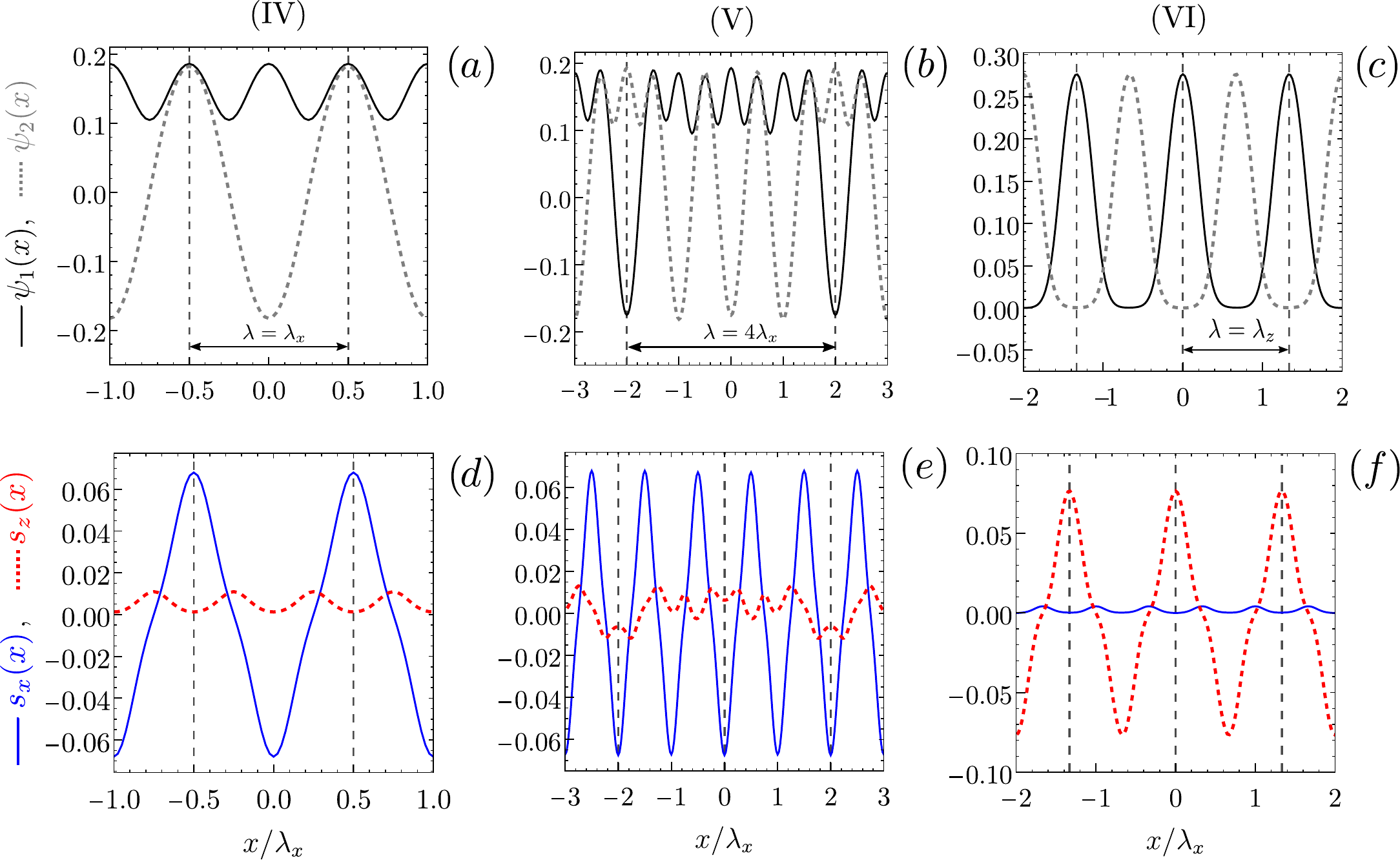}
    \caption{
     Steady-state wavefunctions and pseudo-spin components corresponding to the magnetic configurations indicated in \ref{fig_slices}. The competing orders manifest as a sharp transition between the $\mathcal{M}_{x,\pi}$ and $\mathcal{M}_{z,\pi}$ AFM orders, as shown in $(a)$, $(d)$ and , $(c)$, $(f)$ respectively. The panels $(b)$ and $(e)$ shows the existence of a transient steady state of extended periodicity, but within a narrower interval of $J_z$ values compared with the coexistence region of Fig. \ref{fig_slices}$(a)$, Fig. \ref{fig_wavefunctions1-3}$(b)$ and Fig. \ref{fig_wavefunctions1-3}$(c)$. 
   }
    \label{fig_wavefunctions4-6}
\end{figure}
\noindent
One important feature of the self-organization order parameter diagrams is the existence of regions where both order parameters coexist leading to mixed configurations. The physical effects of the coexistence orders are shown in Fig. \ref{fig_wavefunctions1-3} and Fig. \ref{fig_wavefunctions4-6}, where the wavefunctions corresponding to the order parameters indicated in Fig. \ref{fig_slices} are depicted. Interestingly, the resulting BEC configuration acquires a concrete wavelength that remains ``locked" when both spatial order parameters are nonzero, as shown in Fig. \ref{fig_wavefunctions1-3} (b) and Fig. \ref{fig_wavefunctions1-3} (c). Moreover, we  observe that in the coexistence regime there are spatial oscillations of wavelength  $\lambda_x/2$ inherited from the $\mathcal{M}_{x,\pi}$ super-radiant phase that are consistent with the analysis of the semiclassical energy functional. Our numerical explorations show that the wavelength of the BEC components arising from the self-organization order coexistence is $\lambda = n\lambda_x = m\lambda_z$ when $\xi_{zx}=m/n$ is an irreducible rational number.
This is the fundamental reason why the BEC configurations in Fig. \ref{fig_wavefunctions1-3} (c) and Fig. \ref{fig_wavefunctions4-6} (b) exhibit the same wavelength, even though these cases correspond to different values of the parameter $\xi_{zx}$. Mathematically, the periodicity of the emergent effective potential is the minimum common multiple between $\lambda_x$ and $\lambda_z$. A heuristic argument that explains these results can be presented by analyzing the conditions for which both order parameters in equation (\ref{eq_orderParameters}) are nonzero: Under the one mode approximation, the parameter $\mathcal{M}_{x, \pi}$ supports density configurations satisfying $\sqrt{n_1n_2}\sim | \cos{(k_x x)} |$ while $\mathcal{M}_{z, \pi}$ those with $n_1 - n_2 \sim \cos{(k_z x)}$. Thus, the BEC density is expected as a periodic function with wavelength $\lambda = 2n\lambda_x$ provided $\xi_{zx}=m/n$. In contrast, when $\xi_{zx}$ is set as an irrational number, the wavefunctions resulting from the order competition are non-periodic because $\lambda_x$ and $\lambda_z$ are incommensurate. This formation of non-periodic density patterns resulting from the competition between long-range interactions bears resemblance to the formation of amorphous solids studied in \cite{amorphous}.

\begin{figure}[ht!]
    \centering
    \includegraphics[scale = 0.35]{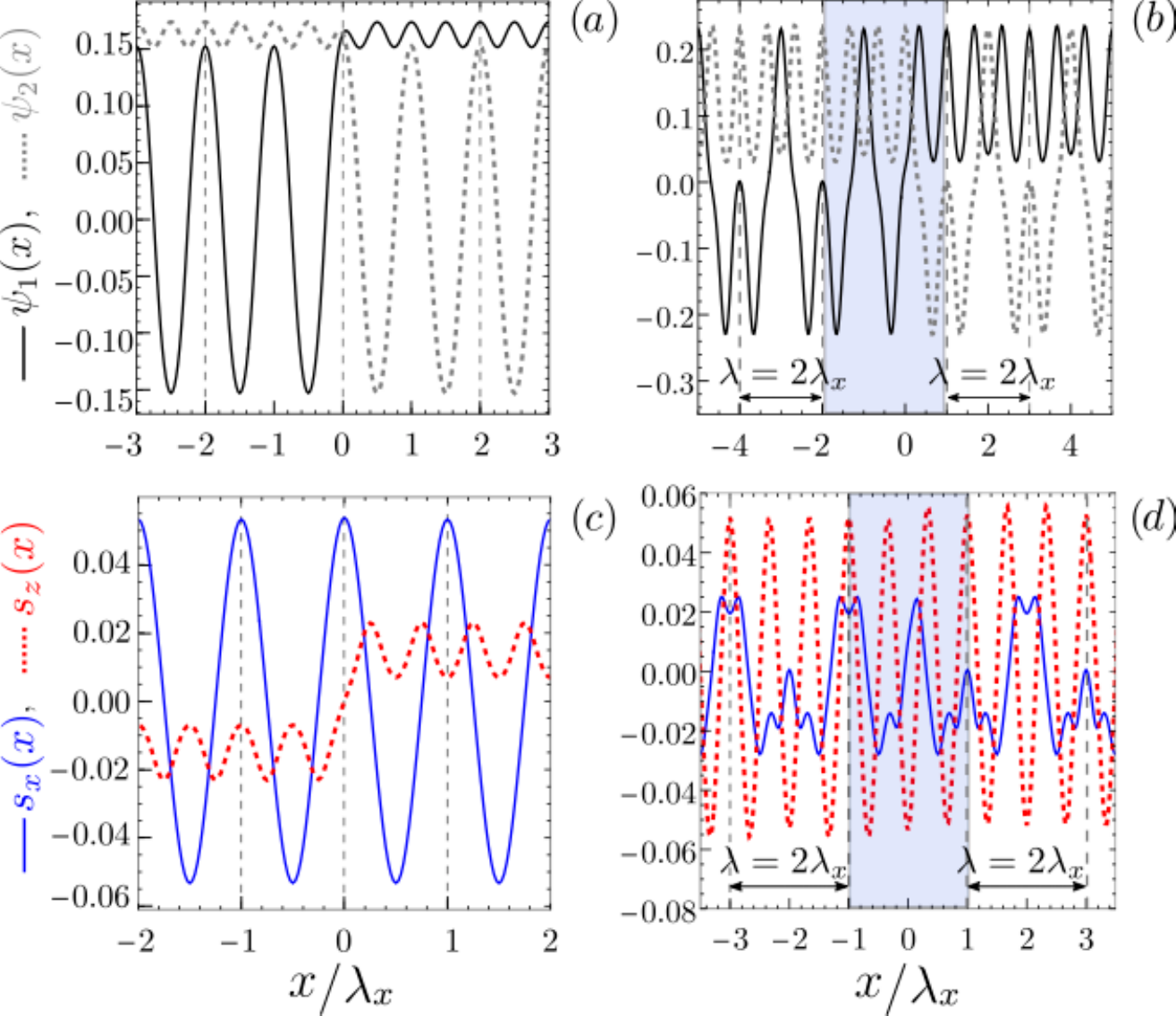}
    \caption{Combined effects of phase separation and self-organization on the BEC wavefunctions. The general phenomena consist of the spatial interchange $\psi_1(x)\leftrightarrow \psi_2(x)$ at the segregation interface, as shown in panels $*a)$ and $(c)$ for the case $J_z=0$. This effect persist even in the intertwine orders $J_\sigma > J^c_\sigma$, as depicted in $(b)$ and $(d)$. The physical parameters are: $(a,c)$ $N(\hbar J_x, Un, U_{12}n)=(3.5, 1.75, 2.0)E_r$. $(b,d)$ $N(\hbar J_x, \hbar J_z, Un, U_{12}n) = (3.0, 5.5, 1.75, 2.0)E_r$ and $\xi_{zx}=3/2$.}
    \label{fig_separationJx}
\end{figure}
\section{Magnetic domain formation in the density segregation regime}\label{sec_separation}
Another degree of experimental control can be accessed by exploring the effects of the density-segregation regime on the self-organization phase transitions. As the previous results rely on the stability analysis of the homogeneous configuration in the mixed regime, it does not allow to make insightful predictions on the density-segregation regime. Nevertheless, the numerical exploration is achievable by the same methods. The key difference is the initial BEC density distribution considered in order to highlight the effects of the phase separation regime on the order competition: first, we computed the steady state for $U_{12}>U$ in absence of optical potentials. Then, we used the phase-separated configuration as seed for the imaginary-time evolution for the general case.
We find that the combined influence of phase separation and spatial self-organization results in segmented configurations, where there is a region from which a spatial exchange of the states of the two BEC components occurs. This phenomena is illustrated in Fig. \ref{fig_separationJx} (a) and Fig. \ref{fig_separationJx} (b) respectively showcasing the wavefunctions and the pseudo-spin components for $J_z =0$, $J_x > J^c_x$ and $U_{12}>U$. In this situation, $x=0$ corresponds to the segmentation interface where the spatial exchange of states occurs. This configuration maintains the pseudo-spin component $s_x(x)$, thus has the same $\mathcal{M}_{x, \pi}$ order parameter as the self-organized state in the mixing regime $U>U_{12}$. It is worth to mention that $s_x(x)$ is invariant under the interchange of states $1\leftrightarrow 2$. Then, the system realize a configuration that locally preserves this symmetry. Additionally, the population imbalance on either side of the interface minimizes the contributions of the two-body interactions to the energy functional. Remarkably, this leads to a ferromagnetic-like ordering in the BEC polarization per atom $s_z(x)$, with opposite signs of its mean value on each side of the segmentation interface.
\begin{figure}[t!]
    \centering
    \includegraphics[scale = 0.15]{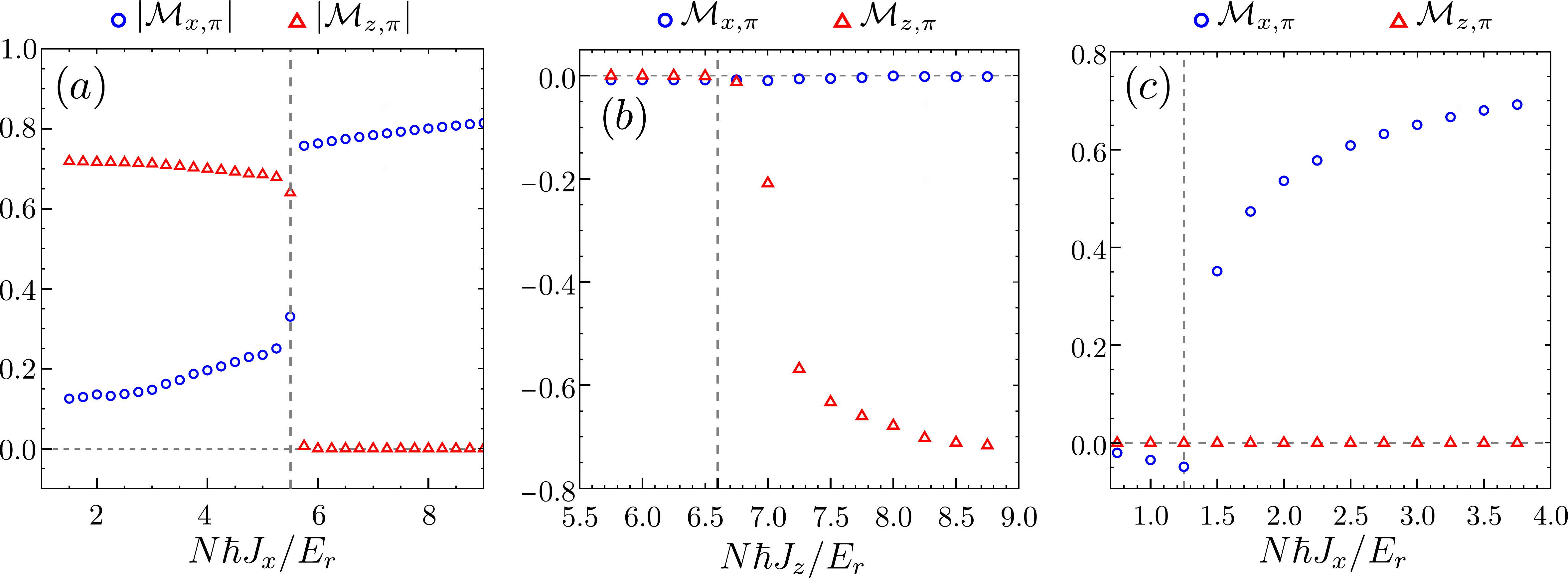}
    \caption{
    AFM order parameters in the density segregation regime. The existence of the competing and coexistence orders (a) and the independent super-radiant transitions (b-c) signal the emergence of self-organization transition in the density segregation regime. The dotted vertical lines in (b-c) denote the self-organization threshold in equations (\ref{eq_Jx_critic_sep}) and (\ref{eq_Jz_critic_sep}) for $N(Un, U_{12}n) = (1.75, 2.0)E_r$.}
    \label{fig_phaseDiagramSeparation}
\end{figure}
This behavior is observed even in the case of coexistence self-organized patterns, as shown in Fig. \ref{fig_separationJx} (c). The order competition stabilizes configurations with extended periodicity $\lambda = n \lambda_x = m \lambda_z$, the separation interface becomes a transition region. Nevertheless, on each side of this region, both the spatial exchange of configurations $1\leftrightarrow 2$ and the emergence of the super lattice with coexistence self-organization orders arise. Interestingly, it is observed that the periodicity $\lambda$ of the wave functions remains robust against short-range two-body atomic interactions. In contrast to the case shown in Fig. \ref{fig_separationJx} (b), the $s_z(x)$ component has antiferromagnetic-like ordering with  periodicity $\lambda$ on both sides of the segregation region.
\\
\noindent
The persistence of the competing self-organization orders in the density segregation regime is elucidated in the behavior of the spatial order parameters. In the general situation $J_\sigma \neq 0$ depicted in Fig. \ref{fig_phaseDiagramSeparation} (a) is observed both, the coexistence ($\mathcal{M}_{\sigma,\pi}\neq 0$) and then the suppression ($\mathcal{M}_{z, \pi} = 0$) of the self-organization configurations as $J_x$ increases. This suggests that the self-organization transition is feasible in the phase segregation regime even if the energy scale associated with the short-range two-body interactions is comparable to the atomic recoil energy, being the self-organization threshold strongly influenced by the atomic dispersion lengths. In order to corroborate this, we compute the spatial order parameters for each self-organization mode independently and report the results in Fig. \ref{fig_phaseDiagramSeparation} (b) and Fig. \ref{fig_phaseDiagramSeparation} (c). Here, the super-radiant transition signaling the self-organization transition is clear.
\\
An estimation of the self-organization threshold in the density segregation regime was obtained by the stability analysis of the initial conditions $\psi_1(x) = \sqrt{n}\theta(x)$ and $\psi_2(x) = \sqrt{n}\theta(-x)$, where $\theta(x)$ is the unit step function. This configuration mimics the steady state in the phase separation regime at position values other than the separation boundary in $x = 0$. At this point, the separation boundary gives rise to a spatial bi-partition of the system generating a homogeneous condensate on each side. Consequently, the stability analysis procedure outlined in section \ref{sec_stability} become applicable. For the region $x<0$, the stability matrix is given by:
\begin{equation}
   \mathcal{L}_{left} = \left( \begin{array}{cccc}
    E_k + V_z & V_z & V_{zx} & V_{zx} \\
    -V_z & -(E_k + V_z) & -V_{zx} & -V_{zx} \\
    V_{xz} & V_{xz} & E_k + V_x & V_{xz} \\
    -V_{xz} & -V_{xz} & -V_{xz} & -(E_k + V_x) 
    \end{array} \right),
    \label{eq_stabilityMatrixLeft}
\end{equation}
with:
\begin{equation}
      \eqalign{
        V_x/N &= (U_{12} - U)n - \delta_{k, \pm k_x}\frac{\hbar J_x}{2},\\
        V_z/N &= Un - \delta_{k, \pm k_z}\frac{\hbar J_z}{2},\\
        V_{xz}/N &= -\delta_{k, \pm k_x}\frac{\hbar J_x}{2},\\
        V_{zx}/N &= \delta_{k, \pm k_z}\frac{\hbar J_z}{2}.
    }
\end{equation}
\begin{figure}[t!]
    \centering
    \includegraphics[scale = 0.35]{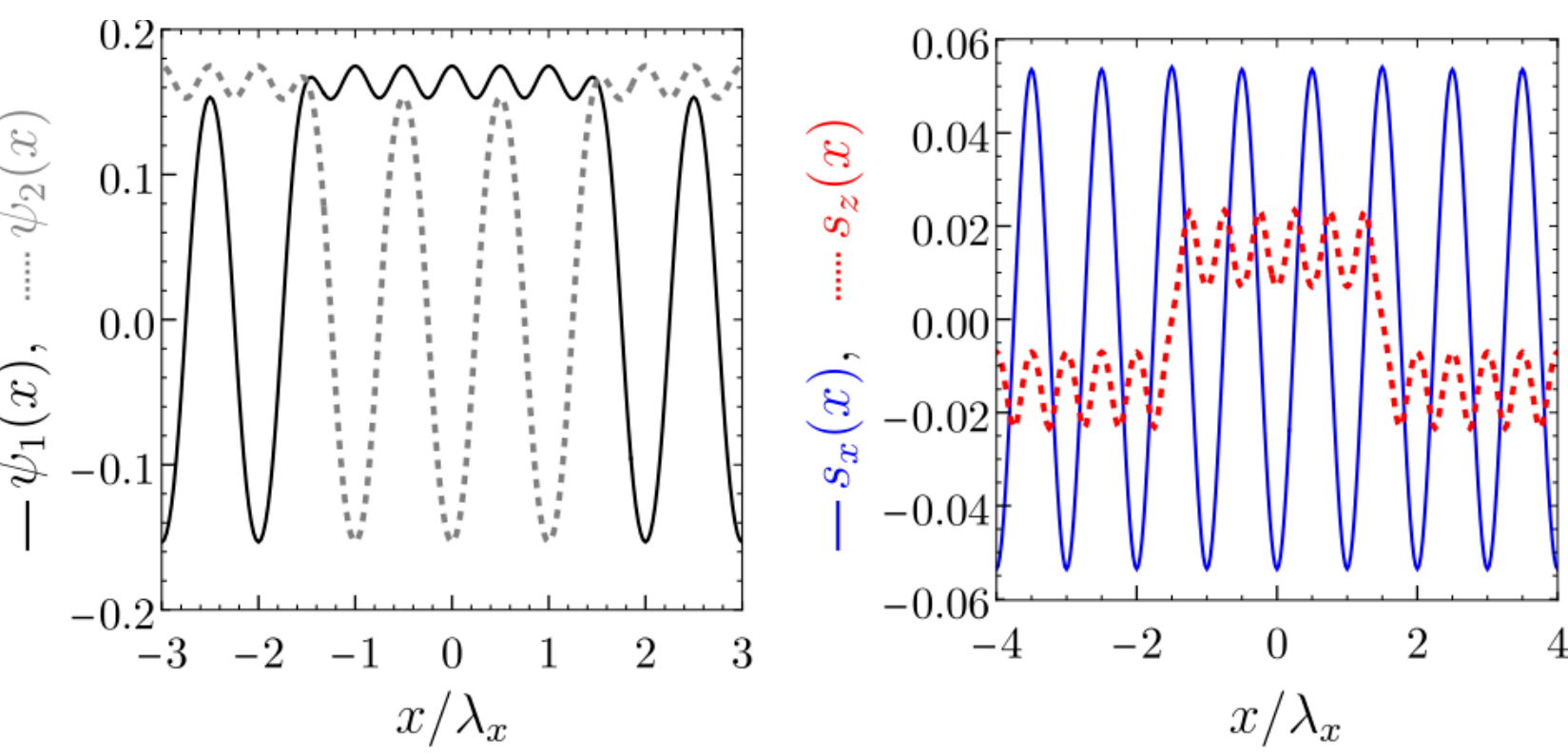}
    \caption{Example of three FM domains generated by the interplay between spatial self-organization and density segregation. The spatial interchange of BEC components generates domains of modulated $\mathcal{M}_{z,0}$ FM orders with opposite magnetization. The physical parameters are those used in Fig. \ref{fig_separationJx}.}
    \label{fig_threeDomain}
\end{figure}
The instability of the mentioned configuration, originated in the exponential growth of density fluctuations, defines the self-organization thresholds given by:
\begin{equation}
    N\hbar J^c_{x} = E_r + N(U_{12} - U)n, 
    \label{eq_Jx_critic_sep}
\end{equation}
\begin{equation}
   N\hbar J^c_{z} = \xi^2_{zx}E^{\phantom{2}}_r + 2NUn,
   \label{eq_Jz_critic_sep}
\end{equation}
which is in good agreement with the numerical results obtained in Fig. \ref{fig_phaseDiagramSeparation}. For the spatial region $x > 0$ the same self-organization thresholds is obtained as consequence of the inversion symmetry of system respect $x=0$.  It follows that, seeding phase separated configurations,  it is possible to simulate  magnetic domain dynamics, such as vanishing  of domains or enhancement of local magnetic response, by manipulating short range interactions. {As depicted in Fig. \ref{fig_threeDomain}, the interplay between phase separation and self-organization provides a mechanism to tailor on-demand FM domains with opposite magnetization values. Conceptually, this arrangement resembles a spin defect in a ferromagnetic material, providing an opportunity to analyze the impact of magnetic impurities on the global magnetic order caused by magnetic impurities \cite{impurities1, Lee2025, RosarioP2023}.}

\section{Conclusions and outlook} \label{sec_conclusions}

Our work explores the self-organization order competition emerging in the steady state of a two component BEC subject to both short- and long-range interactions. The objective of this research was to examine the influence of atom-atom interactions on the phenomenology of cavity-assisted self-organization with the purpose of analog magnetic simulation. In order to achieve this objective, we proposed a framework consisting of a two-component BEC confined within the optical potentials generated by two crossed cavities, each pumped by laser fields. By the adiabatic elimination of the cavity fields, we formulated a description of the BEC dynamics governed by a system of two coupled Gross-Pitaevskii equations with non-local coupling terms. We find our system can support embedded magnetic structures.
\\
Our results illustrate that the two-body interactions have a significant impact in the critical values of the effective Rabi couplings for which the cavity-assisted self-organization occurs: the interplay between short and long-range interactions determines the competition or coexistence of self-organization order parameters akin to magnetic ordering with full parametric control. We have found the existence of stable periodic configurations with an emerging wavelength that depends only on the ratio of the cavity mode wavevectors. Effectively, this implies that the spatial structure of magnetic degrees of freedom can be fully manipulated and controlled parametrically.  Interestingly, when this ratio becomes irrational, the system also supports non-periodic density and magnetic modulations. Our proposal opens the possibility to extend the previous research on optical bistability induced by the transition between localized and extended states \cite{Roati2008, opticalBistability, cavityLocalization} by considering incommensurate optical potentials generated by the coexistence of self-organization orders.
\\
{Experimentally, the preparation of the elongated BEC configuration could be realized by the tight confinement of atoms with a prolate harmonic trap. Integrating the transversal density profile, the effective two-body interaction strength are related with the 3D scattering length via $U_{ij} = 2\hbar\omega_{\bot}a_{ij}$ \cite{Petrov2000}. In addition, the regimes of miscibility and density-segregation of the two-component BEC are accessible by adjusting the s-wave scattering length through a Feshbach resonance. For a variety of known Feshbach resonances, (see Table \ref{Table} and Fig. \ref{fig_magneticField}), the variability on the external magnetic field $B$ necessary to tune $U_{12}/U_{11}$ within the range used in the simulations spans from $0.1$G to $100$G, which is accessible for state of the art experiments in cavity BEC \cite{88Sr(2024)}, where the interplay between long-range interactions, dissipation and internal degrees of freedom enables the characterization and control of effective magnetic orders.}
\\
We suggest that future explorations could delve into the density segregation regime related physics, where potentially impurity regions could accommodate other atoms of a different species or a minority of atoms in an additional spin projection. In the $^{39}$K system perhaps $^{40}$K (fermionic) or $^{41}$K (bosonic) would allow to study impurity physics \cite{Kondo, Anderson} in combinations with long range interactions mediated by the cavity and with different inherent statistical character.  Additionally, extensions in order to consider more spatial dimensions and the dynamical effects resulting from quenching protocols would improve the understanding of the physics of competing and coexistence of orders in many body cavity-quantum electrodynamics. 

\ack
We thank  R. Rosa-Medina, P. Christodoulou, T. Donner, F. Mivehvar and H. Ritsch for helpful discussions. This work is partially  supported by the grants DGAPA-UNAM-PAPIIT: IN118823 and CONAHCYT-CB:A1-S-30934. 
B. R\'\i os-Sanchez acknowledges scholarship from CONAHCYT. 

\appendix
\section{Numerical method}
The results presented in this work have been obtained from the steady state computation of equations (\ref{eq_GPE1}) and (\ref{eq_GPE2}) (GPEs) using the imaginary time evolution method. Initially, we identified $\lambda_x$ and $E_r= \hbar^2 k^2_x /2m =\hbar\omega_r$ as the natural scale of length and energy in order to introduce the dimensionless parameters $\tilde{x}=x/\lambda_x $ and $\tilde{t}=\omega_r t$. Then, the  GPEs are presented in spinor notation as $ \rmi\partial_{\tilde{t}} \Psi(\tilde{x}, \tilde{t}) = (T + V)\Psi(\tilde{x}, \tilde{t})$. For a given initial configuration $\Psi(\tilde{x},\tilde{t}_0)=(\psi_1(\tilde{x}), \psi_2(\tilde{x}))^T$, the imaginary time evolution in a finite step $\Delta \tau$ was calculated by the following split-step scheme:
\begin{equation}
    \eqalign{
        \Psi_{aux1} &= \exp(-\Delta\tau V/2)\Psi(\tilde{x},\tilde{t}_0), \\
        \Psi_{aux2} &= \exp(-\Delta\tau T)\Psi_{aux1}, \\
        \Psi_{aux3} & =\exp(-\Delta\tau V/2)\Psi_{aux2}, \\
        \Psi(\tilde{x}, \tilde{t}_0 + \Delta\tau) & = \Psi_{aux3}\bigg(\int\Psi^\dagger_{aux3}\cdot \Psi_{aux3}\rmd\tilde{x}\bigg)^{-1/2}.       
    }
\end{equation}
In this expressions, the operator $\exp(-\Delta\tau V/2)$ is calculated analytically using the properties of the Pauli matrices. Indeed, for any matrix expressed as a linear combination of Pauli matrices, $A = d_0\sigma_0 + \mathbf{d}\cdot\boldsymbol{\sigma}$, the follow identity holds:
\begin{equation}
    \rme^{A} = \rme^{d_{0}}\bigg[\cosh{(|\mathbf{d}|)}\sigma_0 + \frac{\sinh{(|\mathbf{d}|)}}{|\mathbf{d} |}\mathbf{d}\cdot\boldsymbol{\sigma} \bigg].
\end{equation}
Additionally, $\Psi_{aux2}$ can be computed efficiently using the Fast Fourier Transform algorithm. This routine is iterated until the desired convergence of the spatial order parameters is achieved. The stability of the resulting configuration was tested by implementing the real time evolution algorithm ensuring that the norm of each BEC wavefunction remains constant during the propagation. For our simulations we used a spatial interval of longitude $L_x/\lambda_x=26$, divided by $N_x=1025$ points and a imaginary-time step of $\Delta\tau = 10^{-4} $.

\section{Details on the effective model}
\label{ap:modelDetails}
In order to provide a concrete ground to the physics we intend to explore, let us comment on the relevant approximations and considerations underlying the effective model presented in section \ref{sec_model}:
\begin{itemize}
	{
	\item \textit{Rotating frame:} Under the dipolar and rotating wave approximations, the one-body Hamiltonian describing the dynamics of one atom in the system considered in Section \ref{sec_model} is given by $\hat{h} = \hat{h}_0 + \hat{h}^{\mathrm{AC}}_1 + \hat{h}^{\mathrm{AC}}_2 $, with
	\begin{equation}
		\hat{h}_0 = \frac{\hat{p}^2}{2m} + \sum_\mu \hbar\omega_{\mu\mu}\hat{\sigma}_{\mu\mu} + \sum_i \hbar\omega_{ci}\hat{a}^\dagger_i \hat{a}_i,
	\end{equation}
	\begin{equation}
		\hspace{-1.25cm}\hat{h}^{\mathrm{AC}}_1 = \hbar\big(g_1(x)\hat{a}_1 + \Omega_{b2}\rme^{-i \omega_ {p2}t}\big)\hat{\sigma}_{b2}+  \hbar\big(g_1(x)\hat{a}_1 + \Omega_{a1}\rme^{-i \omega_ {p1}t}\big)\hat{\sigma}_{a1} + \mathrm{H.c.}  
	\end{equation}
	\begin{equation}
		\hspace{-1.25cm}\hat{h}^{\mathrm{AC}}_2 = \hbar g_2(x)\hat{a}_2\big(\hat{\sigma}_{d2} + \hat{\sigma}_{c1} \big) + \hbar\Omega_{c2}\rme^{-i \omega_ {p4}t}\hat{\sigma}_{c2} + \hbar\Omega_{d1}\rme^{-i \omega_ {p3}t}\hat{\sigma}_{d1}+\mathrm{H.c.}
	\end{equation}
	Here, $\hat{\sigma}_{ij} = \ket{i}\bra{j}$ are the atomic transition operators, $m$ is the reduced atomic mass, $\omega_\mu$ and $\omega_{ci}$ are the atom and cavity bare frequencies and $\omega_{\mu j}$ corresponds to the pumping laser frequency tuning the $\ket{j}\rightarrow \ket{\mu}$ transition. Using the methods in \cite{barnett2002methods}, it is possible to construct the following rotation $\hat{h^{\prime}} =\hat{U}\hat{h}\hat{U^\dagger} + i\hbar(\partial_t \hat{U})\hat{U}^\dagger$, where 
	\begin{equation}
		\hspace{-1.75 cm}\eqalign{\hat{U} = & \mathrm{exp}\bigg\{i\bigg[ \omega_{p2}\hat{a}^\dagger_1 \hat{a}_1 + \bigg(\frac{\omega_{p4} + \omega_{p3}}{2}\bigg)\hat{a}^\dagger_2 \hat{a}_2 + \bigg(\frac{\omega_{p4} - \omega_{p3}}{2}\bigg)\hat{\sigma}_{11} + \omega_{p4}\hat{\sigma}_{cc}
		\bigg]t \bigg\} \times\\
		&\mathrm{exp}\bigg\{i\bigg[\bigg(\frac{\omega_{p4} + \omega_{p3}}{2}\bigg)\hat{\sigma}_{dd} + \bigg(\frac{\omega_{p4} - \omega_{p3}}{2} + \omega_{p1}\bigg)\hat{\sigma}_{aa} + \omega_{p2}\hat{\sigma}_{bb} \bigg]t \bigg\},
	}
	\end{equation}
	such that in the limit $\omega_{p1} = \omega_{p2}$ renders the Hamiltonian in a time independent form. Due to the second term in $\hat{h^{\prime}}$, the bare atom and cavity frequencies are shifted as
	\begin{equation}
		\hspace{-2.5 cm}\eqalign{&\hat{h}_0 \rightarrow  \frac{\hat{p}^2}{2m} - \hbar(\omega_{p2} - \omega_{c1})\hat{a}^\dagger_1 \hat{a}_1 - \hbar\bigg(\frac{\omega_{p4} + \omega_{p3}}{2} - \omega_{c2}\bigg)\hat{a}^\dagger_2 \hat{a}_2 + \hbar \omega_{22} \hat{\sigma}_{22}\\
		&-\hbar\bigg(\frac{\omega_{p4} - \omega_{p3}}{2}-\omega_{11}\bigg)\hat{\sigma}_{11} - \hbar\bigg(\frac{\omega_{p4} - \omega_{p3}}{2} + \omega_{p1}-\omega_{aa}\bigg)\hat{\sigma}_{aa} - \hbar(\omega_{p2}-\omega_{bb})\hat{\sigma}_{bb}\\
		&-\hbar(\omega_{p4}-\omega_{cc})\hat{\sigma}_{cc}-\hbar\bigg(\frac{\omega_{p4} + \omega_{p3}}{2}-\omega_{dd}\bigg)\hat{\sigma}_{dd} = \frac{\hat{p}^2}{2m} - \sum_\mu \hbar\Delta_{\mu}\hat{\sigma}_{\mu\mu} -\sum_i \hbar\Delta_{ci}\hat{a}^\dagger_i \hat{a}_i.
	}
	\end{equation}
}
	
	\item \textit{Adiabatic elimination of the excited subspace:} In the far and red-detuning limit, the quantities $1/\Delta_j$ (for $j = a, b, c, d$) represent the fast time scales. Therefore, the steady state configurations $\hat{\Psi}^{ss}_{j}$ 
    in the excited subspace set in a fast transient time and a low atomic occupation. On addition, the kinetic energies $\sim \partial^2_x \hat{\Psi}_j$ can be neglected in comparison with $\hbar\Delta_j$ \cite{opticalPotential2008, reviewcQED}. This allows to eliminate the dynamics of the excited subspace by considering $\hat{\Psi}_j \approx \hat{\Psi}^{ss}_j$, where:
    \begin{equation}
    \hat{\Psi}^{ss}_{a/b} \approx \frac{1}{\Delta_{a/b}}\bigg(g_1(x)\hat{a}_1  + \Omega_{a1/b2}\bigg)\hat{\Psi}_{1/2},
    \label{eq_psi_ab_steady}
    \end{equation}
    \begin{equation}
     \hat{\Psi}^{ss}_{c/d} \approx \frac{1}{\Delta_{c/d}}\bigg(g_2(x)\hat{a}_2\hat{\Psi}_{1/2} + \Omega_{c2/d1}\hat{\Psi}_{2/1}\bigg).
     \label{eq_psi_cd_steady}
    \end{equation}

    \item \textit{Detailed balance of Rabi and laser frequencies:} Following a similar argument, the adiabatic elimination of the cavity field operators with decay rates $\kappa_i$ from the Heisenberg-Langevin equations results in the following expectation values in the steady state:
    \begin{equation}
    \alpha_z=\langle\hat{a}^{ss}_1\rangle \approx \frac{1}{\Delta^{\mathrm{eff}}_{1}} \int \rmd x g_1(x)\bigg[\frac{\Omega_{a1}}{\Delta_{a}}\langle\hat{\Psi}^\dagger_1\hat{\Psi}^{\phantom{\dagger}}_1 \rangle + \frac{\Omega_{b2}}{\Delta_{b}}\langle\hat{\Psi}^\dagger_2\hat{\Psi}^{\phantom{\dagger}}_2 \rangle \bigg].
    \label{eq_az_steady}
    \end{equation}
    
    \begin{equation}
    \alpha_x=\langle\hat{a}^{ss}_2 \rangle\approx \frac{1}{\Delta^{\mathrm{eff}}_{2}} \int \rmd x g_2(x)\bigg[\frac{\Omega_{c2}}{\Delta_{c}}\langle\hat{\Psi}^\dagger_1\hat{\Psi}^{\phantom{\dagger}}_2\rangle + \frac{\Omega_{d1}}{\Delta_{d}}\langle\hat{\Psi}^\dagger_2\hat{\Psi}^{\phantom{\dagger}}_1\rangle\bigg],
    \label{eq_ax_steady}
    \end{equation}
    with:
    \begin{eqnarray}
        {\Delta^{\mathrm{eff}}_{1/2}}& =& \tilde{\Delta}_{1/2} + i\kappa_{1/2},
        \label{eq_cavityDetunings}
        \\
		\tilde{\Delta}_{1/2}&=&\Delta_{c1/c2} - \int \rmd x g^2_{1/2}(x)\bigg(\frac{\langle \hat{\Psi}^\dagger_1\hat{\Psi}^{\phantom{\dagger}}_1\rangle}{\Delta_{c/a}} + \frac{\langle \hat{\Psi}^\dagger_2\hat{\Psi}^{\phantom{\dagger}}_2 \rangle}{\Delta_{d/b}} \bigg),
	\label{disps}
    \end{eqnarray}
    where the last term in (\ref{disps}) is the cavity dispersive shift. 
    We adopt the balanced condition $\Omega_{a1}\Delta_b = - \Omega_{b2}\Delta_a$ in order to couple the amplitude field $\alpha_z$ with the density polarization $\hat{S}_z = \hat{\Psi}^\dagger_1\hat{\Psi}^{\phantom{\dagger}}_1 - \hat{\Psi}^\dagger_2\hat{\Psi}^{\phantom{\dagger}}_2$. With regard of the experimental implementation, the condition $\Delta_{a1} \sim - \Delta_{b2}$ implies that one of the optical transitions is blue detuned respect to the atomic energy splitting. 
    {On the other hand, the condition $\Omega_{a1} \sim -\Omega_{b2}$ can be realized by means of choosing a relative angle of $\pi$ between the polarization vector of the two pump laser fields.} Similarly, we set $\Omega_{c2}\Delta_d =  \Omega_{d1}\Delta_c$ in order to couple the amplitude field $\alpha_x$ with the pseudo-spin polarization $\hat{S}_x = \hat{\Psi}^\dagger_1\hat{\Psi}^{\phantom{\dagger}}_2 + \hat{\Psi}^\dagger_2\hat{\Psi}^{\phantom{\dagger}}_1$. The balance of the Rabi frequencies allows the identification of the following effective Rabi frequencies:
    \begin{equation}
         J_{z} = -\frac{2\tilde{\Delta}_{1}\tilde{g}^2_{1}}{\tilde{\Delta}_{1}^2+\kappa_{1}^2}, \quad  J_{x} = -\frac{2\tilde{\Delta}_{2}\tilde{g}^2_{2}}{\tilde{\Delta}_{2}^2+\kappa_{2}^2},
         \label{js}
    \end{equation}
 with $\tilde{g}_2=g_{02}\Omega_{c2}/\Delta_c$ and  $\tilde{g}_1={g_{01}\Omega_{a1}}/{\Delta_a}$. The value of $\hbar J_{\sigma}$ ($\sigma = x, z$) define the energy scale of the dynamically-generated optical potentials in each cavity.

    \item \textit{Weak coupling limit:} We neglect the dispersive shift, which is approximately constant,  in the effective cavity frequencies $\Delta^{\mathrm{eff}}_i$ in equation (\ref{eq_cavityDetunings}) by considering $Ng^2_{02} \ll J_{x}\Delta_c$ and $Ng^2_{01} \ll J_{z}\Delta_a$.

\end{itemize}

\begin{table}[h]
\caption{\label{Table}Parameters of Feshbach resonances for selected atomic systems. Here is characterized the quantum numbers of the income scattering channel $(F_1, F_2)$ and $m_{F_1}, m_{F_2}$, the magnetic field $B_0$ at which the resonance takes place, the width of the resonance $\Delta$ and the backgrounds scattering length $a_{bg}$ in units of Bohr radius $a_0$. The not founded values are denoted by N.F.}
\begin{indented}
\item[]\begin{tabular}{@{}lllllllll}
\br
\textrm{Atom}& ij&$(F_1, F_2)$ & $m_{F_1}, m_{F_2}$ & $B^{ij}_0$ (\textrm{G}) & $\Delta^{ij}$ (G) & $a^{ij}_{bg} (a_0)$ &  \textrm{Ref.}\\
\mr
\textsuperscript{39}K& 22&(1, 1)& $0, 0$ & 66.0 & -5.403 &-21.09 &  \cite{39K(2010)}  \\
\textsuperscript{39}K& 11&(1, 1)& $+1, +1$ & 25.85 & -0.43 &-35.73 &  \cite{39K(2010)} \\
\textsuperscript{39}K& 12&(1, 1)& $+1, 0$ & 25.8 & -1.25 &-38.02 & \cite{39K(2010)}  \\
\mr
\textsuperscript{7}Li&11 &(1, 1)& $+1, +1$ & 738.2 & -192.3 &-25.8 &  \cite{7Li(2019)}\\
\textsuperscript{7}Li& 22&(1, 1)& $0, 0$ & 844.9 & -14.9 &-23.0 &  \cite{7Li(2019)}\\
\textsuperscript{7}Li& 12&(1, 1)& $+1, 0$ & 794.59 & -90.5 &-29.8 &  \cite{7Li(2019)}\\
\mr
\textsuperscript{23}Na& 22&(1, 1)& $+1, +1$ & 853.0 & 0.0025 & 63.0 &   \cite{FeshbachData2010} \\
\textsuperscript{87}Rb& 11&(1, 1)& $+1, +1$ & 911.7 & 0.0013 & 100.0 &  \cite{FeshbachData2010}\\
\textsuperscript{23}Na $+$ \textsuperscript{87}Rb & 12&(1, 1)& $+1, +1$ & 478.82 & 3.495 & 71.78   & \cite{FeshbachData2010, NaRb(2022)}\\
\mr
\textsuperscript{39}K& 11&(1, 1)& $0, 0$ & 66.0 & -5.403 &-21.09 & \cite{39K(2010)} \\
\textsuperscript{87}Rb& 22&(1, 1)& $+1, +1$ & 911.7 & 0.0013 & 100.0 &  \cite{FeshbachData2010}\\
\textsuperscript{39}K $+$ \textsuperscript{87}Rb& 12 &(1, 1)& $+1, +1$ & 247.9 & 0.28 & 34 & \cite{39K87Rb(2008)}\\
\mr
\textsuperscript{41}K& 11&(1, 1)& $-1, -1$ & 51.1 & -0.361 & 65.1 &  \cite{41K(2009), PotassiumMixtures} \\
\textsuperscript{41}K&12 &(1, 1)& $-1, 0$ & 51.95 & -0.0978 & 65.1 &   \cite{PotassiumMixtures} \\
\mr
\textsuperscript{39}K& 22&(1, 1)& $0, 0$ & 66.0 & -5.403 &-21.09 &  \cite{39K(2010)} \\
\textsuperscript{41}K& 11&(1, 1)& $-1, -1$ & 51.1 & -0.361 &65.1 & \cite{41K(2009), PotassiumMixtures} \\
\textsuperscript{39}K + \textsuperscript{41}K & 12&(1, 1)& $0, -1$ & 228.88 & 0.989 & 171.8 &  \cite{PotassiumMixtures}\\
\mr
\textsuperscript{133}Cs& 11&(3, 3)& $+3, +3$ & 17.0 & 13.0 &-675.0 &   \cite{cesium(1999)} \\
\textsuperscript{85}Rb + \textsuperscript{133}Cs & 12&(2, 3)& $+2, +3$ & 107.13 & -0.17 & -628.0 &  \cite{85Rb133Cs}\\
\mr
\textsuperscript{170}Yb& 11&(0, 0)& $0, 0$ & N.F & N.F  &63.9 &    \cite{Yb(2013)}\\
\textsuperscript{170}Yb& 12&(0, 2)& $0, +2$ & 360 & 2.1 &-64.2   & \cite{Yb(2013)} \\
\textsuperscript{170}Yb& &(0, 2)& $0, -2$ & 1.12 & 2.1 &119.1 &\cite{Yb(2013)}    \\
\mr
\textsuperscript{23}Na& &(1, 1)& $+1, +1$ & 853.0 & 0.0025 & 63.0   & \cite{FeshbachData2010}\\
\textsuperscript{23}Na& &(1, 1)& $-1, -1$ & 1195.0 & -4.0 & 52.98 &   \cite{23Na(2000), 23Na(1999)}\\
\textsuperscript{23}Na& &(1, 1)& $+1, 0$ & N.F & N.F & 52.98 &   \cite{23Na(2000)}\\
\textsuperscript{23}Na& &(1, 1)& $0, 0$ & N.F & N.F & 51.12 &   \cite{23Na(2000)}\\
\br
\end{tabular}
\end{indented}
\end{table}

\noindent{Finally, we evaluate the experimental feasibility of implementing a concrete atomic system by estimating the external magnetic corresponding to the simulation parameters used to report our results. Explicitly, the relation between $U_{ij}$ and the applied magnetic field $B$ near a Feshbach resonance is given by}
\begin{equation}
    U_{ij} = 2 \pi \hbar \omega_\bot a_{ij}, \hspace{0.5cm} a_{ij} = a^{ij}_{bg}\bigg(1 - \frac{\Delta^{ij}}{B - B^{ij}_0} \bigg).
    \label{eq:resonances}
\end{equation}
Here, $\omega_\bot$ is the transversal harmonic frequency confining the BEC, $a^{ij}_{bg}$ is the background scattering length, $\Delta^{ij}$ is the resonance width and $B^{ij}_0$ is the magnetic field at which the Feshbach resonance occurs. In general, these quantities depend on the scattering channel denoted by $ij$. From equation (\ref{eq:resonances}):
\begin{equation}
    \frac{U_{12}}{U_{11}} = \frac{a^{12}_{bg}}{a^{11}_{bg}}\bigg(1 - \frac{\Delta^{12}}{B - B^{12}_0} \bigg)\bigg(1 - \frac{\Delta^{11}}{B - B^{11}_0} \bigg)^{-1}.
    \label{eq:magneticField}
\end{equation}
In Table (\ref{Table}) we report the physical parameters for a selected atom species used in the estimation of the external magnetic field.
\begin{figure}[t!]
    \centering
    \includegraphics[scale = 0.25]{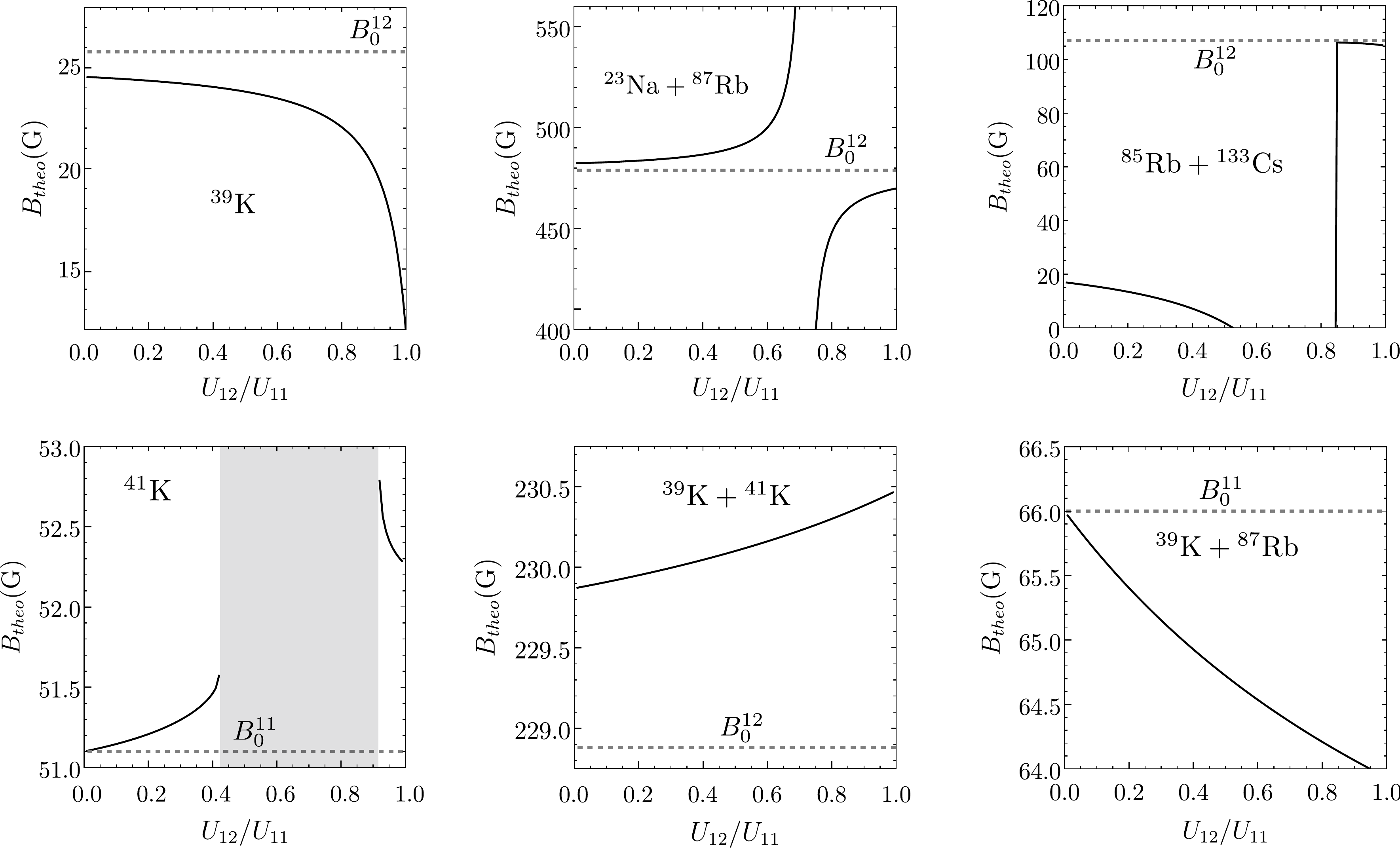}
    \caption{External magnetic field necessary to tune $U_{12} / U_{11} \in [0, 1]$. The shaded region indicates the region where the equation \ref{eq:magneticField} does not have a real solution for $B$, indicating the limitations of the simply model of Feshbach resonance considered.}
    \label{fig_magneticField}
\end{figure}

\section{Spatial frequency behaviour and coexistence region}
The coexistence of AFM orders shown in Fig. \ref{fig_slices}(a), \ref{fig_wavefunctions1-3}(b) and Fig. \ref{fig_wavefunctions1-3}(c) is also appreciated in the distribution of the spatial frequencies of BEC wavefunctions as $J_x$ increases. Indeed, from the initial AFM order $\mathcal{M}_{z,\pi}$ with a peak in $k=k_z$ emerges a configuration with a momentum distribution peaked in $k=2k_x - k_z$, $k=k_x/2$, $k=k_z - k_x$ and $k=k_x$ which induce BEC wavefunctions with extended periodicity $\lambda = 4\lambda_x$.
\begin{figure}[h!]
    \centering    \includegraphics[scale = 0.2]{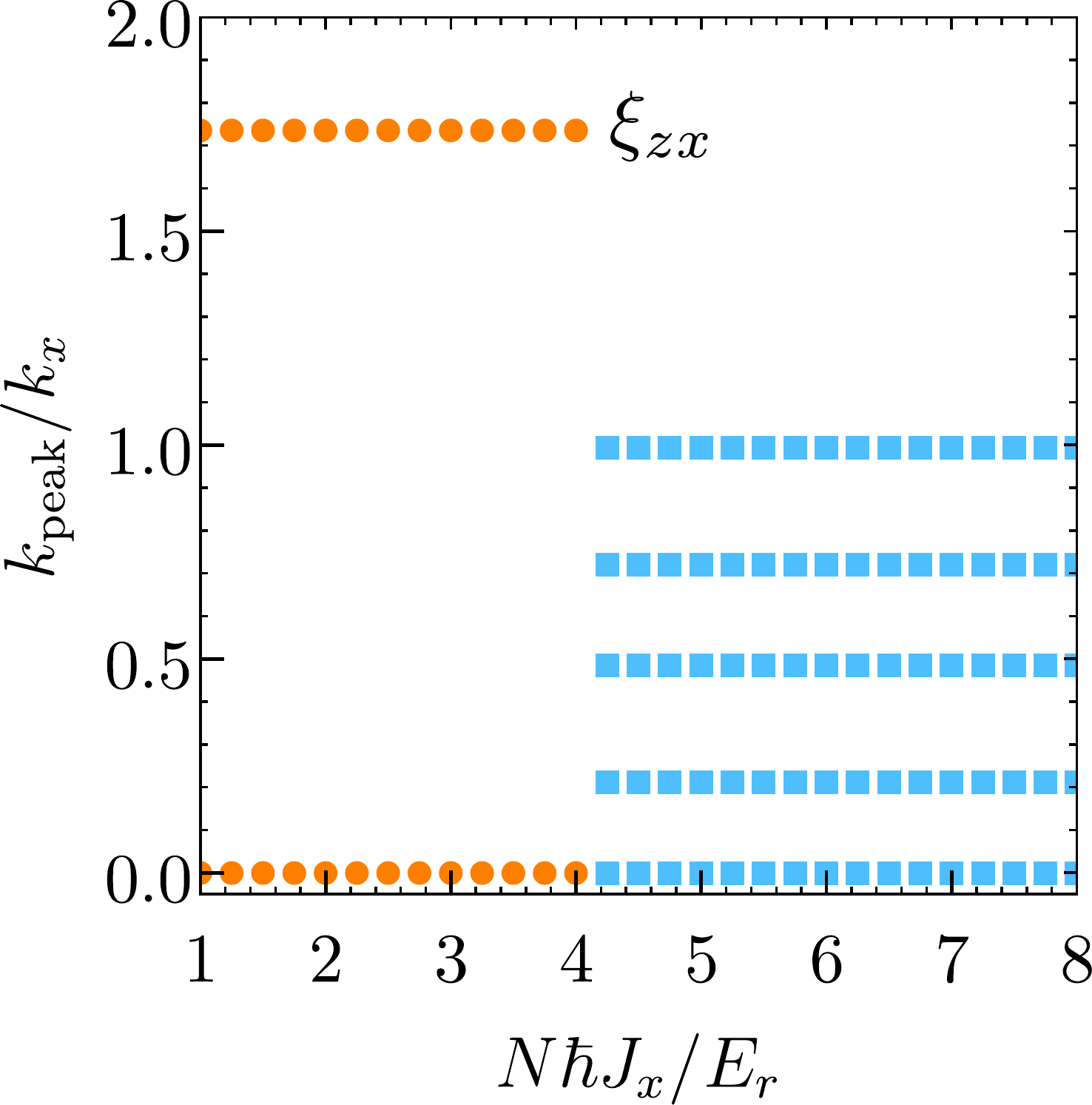}
    \caption{Momentum peak distribution for the transition to the coexistence AFM order in Fig. \ref{fig_slices}(a). For $N\hbar J_x > 4 E_r $ the BEC wavefunctions exhibit momentum components lesser than $k_z = \xi_{zx}k_x$ that indicates the emergence of the extended-periodic density configuration. Here $\xi_{zx}=7/4.$
    }
    \label{fig_momentum}
\end{figure}

\section{$\mathcal{M}_{x, \pi}$ phase under the two mode approximation.}
\label{sec_appendix_twoMode}
As stated in the main text, the $\mathcal{M}_{x, \pi}$ AFM order is characterized by the energy locking of the relative phase between the BEC wavefunctions such that $\cos{\big(\arg(\psi_1) - \arg(\psi_2) \big)} = \pm 1$. For real-valued wavefunctions, the two-mode approximation consistent with this condition is
\begin{equation}
	\hspace{-2cm}
    \psi_1 = \sqrt{\frac{2}{L_x}}
    c_1\cos{(k_x x)}, \hspace{0.5cm} \psi_2 = \sqrt{\frac{1}{L_x}}\big(c_0 + \sqrt{2}c_2\cos{(2 k_x x)}\big), \hspace{0.5cm} c_0^2 + c_1^2 + c_2^2 = 1,
\end{equation}
being $c_i$ the variational parameters. Prior to presenting the expression for the energy functional, let us discuss the impact of commensurability on the integrals of the optical potentials. For odd powers of the cosine function we have:
\begin{equation}
	\hspace{-1cm}
    \frac{1}{L_x}\int \rmd x \cos^{2n+1}{(k_\sigma x)} = \frac{1}{k_\sigma L_x} \sin{(k_\sigma L_x)}\, _2F_1\bigg(\frac{1}{2},-n, \frac{3}{2}; \sin^2{(k_\sigma L_x)} \bigg),
\end{equation}
being $\, _2F_1(a,b,c; z)$ the hypergeometric function. Note that if the potential wavelength $\lambda_\sigma$ and the BEC length are commensurate, that means $L_x/ \lambda_\sigma$ is an integer, this integral is strictly zero. In other cases, this integral can be negligible in comparison with the integrals of the even powers of cosine function as $(k_\sigma L_x)^{-1} \ll 1$. As a consequence, this implies that deep in one of the AFM orders, the remaining one becomes energetically suppressed. Under these considerations, the evaluation of the following integrals:
\begin{equation}
    \mathcal{M}_{z,0} = \int \rmd x (n_1 - n_2) = c_1^2 - (c_0^2 + c_2^2) = 1-2(c_0^2 + c_2^2),
\end{equation}
\begin{equation}
	\hspace{-2cm}
    \mathcal{M}_{x,\pi} = \frac{2\sqrt{2}}{L_x}c_1 \int \rmd x \cos^2{(k_x x)} \big(c_0 + \sqrt{2}c_2\cos{(2 k_x x)} \big) = (1-c_0^2 - c_2^2)^{1/2}(\sqrt{2}c_0 + c_2),
\end{equation}
\begin{equation}
    \int \rmd x n_1^2 = \frac{4}{L_x^2}c_1^4 \int \rmd x \cos^4{(k_x x)} = \frac{3}{2}n(1-c_0^2 - c_2^2)^2,
\end{equation}
\begin{equation}
    \int \rmd x n_2^2 = \frac{1}{L_x^2}\int \rmd x \big(c_0 + \sqrt{2}c_2\cos{(2 k_x x)}\big)^4 = n(c_0^4 + 6 c_0^2 c_2^2 + 3c_2^4 / 2),
\end{equation}
\begin{equation}
    \eqalign{
        \int \rmd x n_1 n_2 &= \frac{2}{L_x^2}c_1^2 \int \rmd x \cos^2{(k_x x)} \big(c_0 + \sqrt{2}c_2\cos{(2 k_x x)}\big)^2 
        \\
        &= n(1-c_0^2 - c_2^2) (c_0^2 + \sqrt{2} c_0 c_2 + c_2^2)
    }    
\end{equation}
with $n = L_x^{-1}$, result in the following energy functional:
\begin{equation}
	\hspace{-1cm}
    \eqalign{
        E_{\mathcal{M}_x} &= 3 E_r c_2^2 - E_r c_0^2 -\hbar \delta(c_0^2 + c_2^2)- \frac{1}{2}N\hbar J_x (1-c_0^2 - c_2^2)(\sqrt{2}c_0 + c_2)^{2}\\
        &+ \frac{3}{4}nNU_{11}(1-c_0^2 - c_2^2)^2 + nNU_{12}(1-c_0^2 - c_2^2) (c_0^2 + \sqrt{2} c_0 c_2 + c_2^2)\\
        &+\frac{1}{2}nNU_{22}\big(c_0^4 + 6 c_0^2 c_2^2 + \frac{3}{2}c_2^4\big),
    }
\end{equation}
where it has been used that $c_1^2=1-c_0^2-c_2^2$.
\section{$\mathcal{M}_{z, \pi}$ phase under the two mode approximation.}
\label{sec_appendix_twoModeMz}
Following the same procedure stated before, for the $\mathcal{M}_{z, \pi}$ AFM order we have:
\begin{equation}
    \mathcal{M}_{z,0} = \int \rmd x (n_1 - n_2) = a_0^2 + a_1^2 - b_0^2 - b_1^2,
\end{equation}

\begin{equation}
    \eqalign{
        \mathcal{M}_{z,\pi} &= \frac{1}{L_x}\int \rmd x\cos{(k_z x)}\big( a_0^2 - b_0^2 + 2\sqrt{2}(a_0a_1 - b_0b_1)\cos{(k_z x)}\\
        &- 2(a_1^2 -b_1^2)\cos^2{(k_z x)}\big)=\sqrt{2}(a_0a_1 -b_0b_1),
    }
\end{equation}

\begin{equation}
    \int \rmd x n_1^2 = \frac{1}{L_x^2}\int \rmd x \big(a_0 + \sqrt{2}a_1\cos{(k_z x)}\big)^4 = n(a_0^4 + 6 a_0^2 a_1^2 + \frac{3}{2}a_1^4),
\end{equation}

\begin{equation}
    \int \rmd x n_2^2 = \frac{1}{L_x^2}\int \rmd x \big(b_0 + \sqrt{2}b_1\cos{(k_z x)}\big)^4 = n(b_0^4 + 6 b_0^2 b_1^2 + \frac{3}{2}b_1^4).
\end{equation}
From these expression the functional energy results in:
\begin{equation}
    \eqalign{
    	\hspace{-1cm}
        E_{\mathcal{M}_z} &= \xi_{zx}^2 E_r (a_1^2 + b_1^2) + \frac{\hbar \delta}{2}\big(a_0^2 + a_1^2 - b_0^2 - b_1^2\big) - N\hbar J_z(a_0 a_1 - b_0 b_1)^2\\
        &+\frac{1}{2}nNU_{11}\big( a_0^4 + 6 a_0^2 a_1^2 + \frac{3}{2}a_1^4\big)+\frac{1}{2}nNU_{22}\big( b_0^4 + 6 b_0^2 b_1^2 + \frac{3}{2}b_1^4\big)\\
        &+nNU_{12}\big(4a_0 a_1 b_0 b_1 + a_0^2(b_0^2 + b_ 1^2)+ a_1^2(b_0^2 + \frac{3}{2}b_ 1^2)\big).
        \label{eq:functionalMz}
    }
\end{equation}
By replacing $b_1 = \sqrt{1 - a_0^2 -a_1^2 - b_0^2}$, the minimization problem becomes unrestricted.

\newpage
\section*{References}

\bibliographystyle{unsrt}
\bibliography{bibliography.bib}

\begin{thebibliography}{10}

\bibitem{quantumSimulator}
Maciej Lewenstein, Anna Sanpera, and Verònica Ahufinger.
\newblock {\em Ultracold Atoms in Optical Lattices: Simulating quantum
  many-body systems}.
\newblock Oxford University Press, 03 2012.

\bibitem{quantumSimulatorPRX}
Ehud Altman, Kenneth~R. Brown, Giuseppe Carleo, Lincoln~D. Carr, Eugene Demler,
  Cheng Chin, Brian DeMarco, Sophia~E. Economou, Mark~A. Eriksson, Kai-Mei~C.
  Fu, Markus Greiner, Kaden~R.A. Hazzard, Randall~G. Hulet, Alicia~J. Koll\'ar,
  Benjamin~L. Lev, Mikhail~D. Lukin, Ruichao Ma, Xiao Mi, Shashank Misra,
  Christopher Monroe, Kater Murch, Zaira Nazario, Kang-Kuen Ni, Andrew~C.
  Potter, Pedram Roushan, Mark Saffman, Monika Schleier-Smith, Irfan Siddiqi,
  Raymond Simmonds, Meenakshi Singh, I.B. Spielman, Kristan Temme, David~S.
  Weiss, Jelena Vu\ifmmode \check{c}\else \v{c}\fi{}kovi\ifmmode~\acute{c}\else
  \'{c}\fi{}, Vladan Vuleti\ifmmode~\acute{c}\else \'{c}\fi{}, Jun Ye, and
  Martin Zwierlein.
\newblock Quantum simulators: Architectures and opportunities.
\newblock {\em PRX Quantum}, 2:017003, Feb 2021.

\bibitem{Cirac2012}
J.~Ignacio Cirac and Peter Zoller.
\newblock Goals and opportunities in quantum simulation.
\newblock {\em Nat. Phys.}, 8(4):264--266, Apr 2012.

\bibitem{Caballero2022}
Karen Lozano-M\'endez, Alejandro~H. Casares, and Santiago~F. Caballero-Benitez.
\newblock Spin entanglement and magnetic competition via long-range
  interactions in spinor quantum optical lattices.
\newblock {\em Phys. Rev. Lett}, 128:080601, 2022.

\bibitem{magneticMoire}
C.~Madro\~nero and R.~Paredes.
\newblock Dynamic stability in spinor bose gases in moir\'e lattices with
  square and hexagonal symmetries.
\newblock {\em Phys. Rev. A}, 107:033316, Mar 2023.

\bibitem{magneticToolbox}
Farokh Mivehvar, Helmut Ritsch, and Francesco Piazza.
\newblock Cavity-quantum-electrodynamical toolbox for quantum magnetism.
\newblock {\em Phys. Rev. Lett.}, 122:113603, Mar 2019.

\bibitem{spinTextures}
M.~Landini, N.~Dogra, K.~Kroeger, L.~Hruby, T.~Donner, and T.~Esslinger.
\newblock Formation of a spin texture in a quantum gas coupled to a cavity.
\newblock {\em Phys. Rev. Lett.}, 120:223602, May 2018.

\bibitem{spinDensityOrder}
Natalia Masalaeva, Wolfgang Niedenzu, Farokh Mivehvar, and Helmut Ritsch.
\newblock Spin and density self-ordering in dynamic polarization gradients
  fields.
\newblock {\em Phys. Rev. Res.}, 3:013173, Feb 2021.

\bibitem{cristallineDipolar}
Chinmayee Mishra, Stefan Ostermann, Farokh Mivehvar, and B.~Prasanna Venkatesh.
\newblock Crystalline phases of laser-driven dipolar bose-einstein condensates.
\newblock {\em Phys. Rev. A}, 107:023312, Feb 2023.

\bibitem{Tricriticality}
Yun Li, Lev~P. Pitaevskii, and Sandro Stringari.
\newblock Quantum tricriticality and phase transitions in spin-orbit coupled
  bose-einstein condensates.
\newblock {\em Phys. Rev. Lett.}, 108:225301, May 2012.

\bibitem{stripeSupersolid}
Jun-Ru Li, Jeongwon Lee, Wujie Huang, Sean Burchesky, Boris Shteynas,
  Furkan~{\c{C}}a{\u{g}}r{\i} Top, Alan~O. Jamison, and Wolfgang Ketterle.
\newblock A stripe phase with supersolid properties in spin--orbit-coupled
  bose--einstein condensates.
\newblock {\em Nature}, 543(7643):91--94, Mar 2017.

\bibitem{timeCrystal1}
Hans Keßler, Jayson~G Cosme, Christoph Georges, Ludwig Mathey, and Andreas
  Hemmerich.
\newblock From a continuous to a discrete time crystal in a dissipative
  atom-cavity system.
\newblock {\em New. J. Phys.}, 22(8):085002, aug 2020.

\bibitem{timeCrystal2}
Hans Ke\ss{}ler, Phatthamon Kongkhambut, Christoph Georges, Ludwig Mathey,
  Jayson~G. Cosme, and Andreas Hemmerich.
\newblock Observation of a dissipative time crystal.
\newblock {\em Phys. Rev. Lett.}, 127:043602, Jul 2021.

\bibitem{LightInducedQuasicrystal}
Farokh Mivehvar, Helmut Ritsch, and Francesco Piazza.
\newblock Emergent quasicrystalline symmetry in light-induced quantum phase
  transitions.
\newblock {\em Phys. Rev. Lett.}, 123:210604, Nov 2019.

\bibitem{cavitySpinOrbit}
Stefan Ostermann, Helmut Ritsch, and Farokh Mivehvar.
\newblock Many-body phases of a planar bose-einstein condensate with
  cavity-induced spin-orbit coupling.
\newblock {\em Phys. Rev. A}, 103:023302, Feb 2021.

\bibitem{spinOrbit}
Y.~Deng, J.~Cheng, H.~Jing, and S.~Yi.
\newblock Bose-einstein condensates with cavity-mediated spin-orbit coupling.
\newblock {\em Phys. Rev. Lett.}, 112:143007, Apr 2014.

\bibitem{phaseSeparationSelforganization}
Abid Ali, Farhan Saif, and Hiroki Saito.
\newblock Phase separation and multistability of a two-component bose-einstein
  condensate in an optical cavity.
\newblock {\em Phys. Rev. A}, 105:063318, Jun 2022.

\bibitem{competingOrders}
Renate Landig, Lorenz Hruby, Nishant Dogra, Manuele Landini, Rafael Mottl,
  Tobias Donner, and Tilman Esslinger.
\newblock Quantum phases from competing short- and long-range interactions in
  an optical lattice.
\newblock {\em Nature}, 532(7600):476--479, Apr 2016.

\bibitem{chiralTopo}
Biao Dong and YongChang Zhang.
\newblock Raman laser induced self-organization with topology in a dipolar
  condensate.
\newblock {\em Opt. Express}, 31(5):7523--7534, Feb 2023.

\bibitem{Morales2018}
Andrea Morales, Philip Zupancic, Julian Leonard, Tilman Esslinger, and Tobias
  Donner.
\newblock Coupling two order parameters in a quantum gas.
\newblock {\em Nat. Mater.}, 17(8):686--690, Aug 2018.

\bibitem{crossedCavities2017}
Julian Leonard, Andrea Morales, Philip Zupancic, Tilman Esslinger, and Tobias
  Donner.
\newblock Supersolid formation in a quantum gas breaking a continuous
  translational symmetry.
\newblock {\em Nature}, 543(7643):87--90, Mar 2017.

\bibitem{PRLSelforganization2017}
Farokh Mivehvar, Francesco Piazza, and Helmut Ritsch.
\newblock Disorder-driven density and spin self-ordering of a bose-einstein
  condensate in a cavity.
\newblock {\em Phys. Rev. Lett.}, 119:063602, Aug 2017.

\bibitem{Higgs}
Julian Leonard, Andrea Morales, Philip Zupancic, Tobias Donner, and Tilman
  Esslinger.
\newblock Monitoring and manipulating higgs and goldstone modes in a supersolid
  quantum gas.
\newblock {\em Science}, 358(6369):1415--1418, 2017.

\bibitem{Zhiqiang:17}
Zhang Zhiqiang, Chern~Hui Lee, Ravi Kumar, K.~J. Arnold, Stuart~J. Masson,
  A.~S. Parkins, and M.~D. Barrett.
\newblock Nonequilibrium phase transition in a spin-1 dicke model.
\newblock {\em Optica}, 4(4):424--429, Apr 2017.

\bibitem{Barret:17}
Stuart~J. Masson, M.~D. Barrett, and Scott Parkins.
\newblock Cavity qed engineering of spin dynamics and squeezing in a spinor
  gas.
\newblock {\em Phys. Rev. Lett.}, 119:213601, Nov 2017.

\bibitem{HelmutRev}
Farokh Mivehvar, Francesco Piazza, Tobias Donner, and Helmut Ritsch.
\newblock Cavity qed with quantum gases: new paradigms in many-body physics.
\newblock {\em Advances in Physics}, 70:1--153, 2021.

\bibitem{HelmutRing}
Stefan Ostermann, Wolfgang Niedenzu, and Helmut Ritsch.
\newblock Unraveling the quantum nature of atomic self-ordering in a ring
  cavity.
\newblock {\em Phys. Rev. Lett.}, 124:033601, 2020.

\bibitem{Caballero2015}
Santiago~F. Caballero-Benitez and Igor~B. Mekhov.
\newblock Quantum optical lattices for emergent many-body phases of ultracold
  atoms.
\newblock {\em Phys. Rev. Lett.}, 115:243604, 2015.

\bibitem{selfbound}
Maria Arazo, Albert Gallem\'{\i}, Montserrat Guilleumas, Ricardo Mayol, and
  Luis Santos.
\newblock Self-bound crystals of antiparallel dipolar mixtures.
\newblock {\em Phys. Rev. Res.}, 5:043038, Oct 2023.

\bibitem{Dalafi_2013}
A~Dalafi, M~H Naderi, M~Soltanolkotabi, and Sh~Barzanjeh.
\newblock Controllability of optical bistability, cooling and entanglement in
  hybrid cavity optomechanical systems by nonlinear atom–atom interaction.
\newblock {\em J. Phys. B: Atom. Mol. Opt. Phys.}, 46(23):235502, nov 2013.

\bibitem{Deng-2023}
Yuangang Deng and Su~Yi.
\newblock Self-ordered supersolid phase beyond dicke superradiance in a ring
  cavity.
\newblock {\em Phys. Rev. Res.}, 5:013002, Jan 2023.

\bibitem{Qin-2022}
Jieli Qin and Lu~Zhou.
\newblock Supersolid gap soliton in a bose-einstein condensate and optical ring
  cavity coupling system.
\newblock {\em Phys. Rev. E}, 105:054214, May 2022.

\bibitem{DrivenSupersolid}
Farokh Mivehvar, Stefan Ostermann, Francesco Piazza, and Helmut Ritsch.
\newblock Driven-dissipative supersolid in a ring cavity.
\newblock {\em Phys. Rev. Lett.}, 120:123601, Mar 2018.

\bibitem{Rogel-Salazar_2013}
J~Rogel-Salazar.
\newblock The gross–pitaevskii equation and bose–einstein condensates.
\newblock {\em European Journal of Physics}, 34(2):247, jan 2013.

\bibitem{selforganization2010}
K~Baumann, Christine Guerlin, Ferdinand Brennecke, and Tilman Esslinger.
\newblock {Dicke quantum phase transition with a superfluid gas in an optical
  cavity}.
\newblock {\em Nature}, 464(7293):1301--1306, April 2010.

\bibitem{pitaevskiiBook}
Lev Pitaevskii and S.~Stringari.
\newblock {\em Bose-Einstein Condensation}.
\newblock International Series of Monographs on Physics. Clarendon Press, 2003.

\bibitem{twoComponentArxiv}
Giacomo Lamporesi.
\newblock Two-component spin mixtures.
\newblock {\em arXiv:2304.03711v2}, 2023.

\bibitem{BogoliubovCoh2003}
Paolo Tommasini, E.~J.~V. de~Passos, A.~F.~R. de~Toledo~Piza, M.~S. Hussein,
  and E.~Timmermans.
\newblock Bogoliubov theory for mutually coherent condensates.
\newblock {\em Phys. Rev. A}, 67:023606, Feb 2003.

\bibitem{excitations}
A~S Alexandrov and V~V Kabanov.
\newblock Excitations and phase segregation in a two-component bose-einstein
  condensate with an arbitrary interaction.
\newblock {\em J. Phys.: Condens. Matter}, 14(18):L327, apr 2002.

\bibitem{abad2013}
Marta Abad and Alessio Recati.
\newblock A study of coherently coupled two-component bose-einstein
  condensates.
\newblock {\em Eur. Phys. J. D}, 67(7):148, Jul 2013.

\bibitem{reviewCoupledBECS}
Alessio Recati and Sandro Stringari.
\newblock Coherently coupled mixtures of ultracold atomic gases.
\newblock {\em Annu. Rev. Condens. Matter Phys.}, 13(1):407--432, 2022.

\bibitem{split-step-1}
L.~Lehtovaara, J.~Toivanen, and J.~Eloranta.
\newblock Solution of time-independent schrödinger equation by the imaginary
  time propagation method.
\newblock {\em J. Comput. Phys.}, 221(1):148--157, 2007.

\bibitem{split-step-2}
Dušan Vudragović, Ivana Vidanović, Antun Balaž, Paulsamy Muruganandam, and
  Sadhan~K. Adhikari.
\newblock C programs for solving the time-dependent gross–pitaevskii equation
  in a fully anisotropic trap.
\newblock {\em Comput. Phys. Commun.}, 183(9):2021--2025, 2012.

\bibitem{orderParameters}
Astrid Eichhorn, David Mesterh\'azy, and Michael~M. Scherer.
\newblock Multicritical behavior in models with two competing order parameters.
\newblock {\em Phys. Rev. E}, 88:042141, Oct 2013.

\bibitem{amorphous}
Stefan Ostermann, Valentin Walther, and Susanne~F. Yelin.
\newblock Superglass formation in an atomic bec with competing long-range
  interactions.
\newblock {\em Phys. Rev. Res.}, 4:023074, Apr 2022.

\bibitem{impurities1}
Dacheng Ma, Yan Qi, and An~Du.
\newblock The exact solution of magnetic susceptibility for finite ising ring
  with a magnetic impurity.
\newblock {\em Canadian Journal of Physics}, 99(11):998--1006, 2021.

\bibitem{Lee2025}
Yu-Li Lee.
\newblock Magnetic impurities in an altermagnetic metal.
\newblock {\em The European Physical Journal B}, 98(3):43, Mar 2025.

\bibitem{RosarioP2023}
C.~Madro\~nero and R.~Paredes.
\newblock Dynamic stability in spinor bose gases in moir\'e lattices with
  square and hexagonal symmetries.
\newblock {\em Phys. Rev. A}, 107:033316, Mar 2023.

\bibitem{Roati2008}
Giacomo Roati, Chiara D'Errico, Leonardo Fallani, Marco Fattori, Chiara Fort,
  Matteo Zaccanti, Giovanni Modugno, Michele Modugno, and Massimo Inguscio.
\newblock Anderson localization of a non-interacting bose--einstein condensate.
\newblock {\em Nature}, 453(7197):895--898, Jun 2008.

\bibitem{opticalBistability}
Subhadeep Gupta, Kevin~L. Moore, Kater~W. Murch, and Dan~M. Stamper-Kurn.
\newblock Cavity nonlinear optics at low photon numbers from collective atomic
  motion.
\newblock {\em Phys. Rev. Lett.}, 99:213601, Nov 2007.

\bibitem{cavityLocalization}
Lu~Zhou, Han Pu, Keye Zhang, Xing-Dong Zhao, and Weiping Zhang.
\newblock Cavity-induced switching between localized and extended states in a
  noninteracting bose-einstein condensate.
\newblock {\em Phys. Rev. A}, 84:043606, Oct 2011.

\bibitem{Petrov2000}
D.~S. Petrov, G.~V. Shlyapnikov, and J.~T.~M. Walraven.
\newblock Regimes of quantum degeneracy in trapped 1d gases.
\newblock {\em Phys. Rev. Lett.}, 85:3745--3749, Oct 2000.

\bibitem{88Sr(2024)}
Eric Song, Diego Barberena, Dylan Young, Edwin Chaparro, Anjun Chu, Sanaa
  Agarwal, Zhijing Niu, Jeremy Young, Ana Rey, and James Thompson.
\newblock A dissipation-induced superradiant transition in a strontium
  cavity-qed system.
\newblock {\em arXiv:2408.11086}, 08 2024.

\bibitem{Kondo}
Jun Kondo.
\newblock {Resistance Minimum in Dilute Magnetic Alloys}.
\newblock {\em Prog. Theo. Phys.}, 32(1):37--49, 07 1964.

\bibitem{Anderson}
P.~W. Anderson.
\newblock Localized magnetic states in metals.
\newblock {\em Phys. Rev.}, 124:41--53, Oct 1961.

\bibitem{barnett2002methods}
Stephen Barnett and Paul~M Radmore.
\newblock {\em Methods in theoretical quantum optics}, volume~15.
\newblock Oxford University Press, 2002.

\bibitem{opticalPotential2008}
Christoph Maschler, Igor~B. Mekhov, and Helmut Ritsch.
\newblock Ultracold atoms in optical lattices generated by quantized light
  fields.
\newblock {\em Eur. Phys. J. D}, 46:545--560, 2007.

\bibitem{reviewcQED}
Farokh Mivehvar, Francesco Piazza, Tobias Donner, and Helmut Ritsch.
\newblock Cavity qed with quantum gases: new paradigms in many-body physics.
\newblock {\em Adv. Phys.}, 70(1):1--153, 2021.

\bibitem{39K(2010)}
M.~Lysebo and L.~Veseth.
\newblock Feshbach resonances and transition rates for cold homonuclear
  collisions between $^{39}\mathrm{K}$ and $^{41}\mathrm{K}$ atoms.
\newblock {\em Phys. Rev. A}, 81:032702, Mar 2010.

\bibitem{7Li(2019)}
Jesse Amato-Grill, Niklas Jepsen, Ivana Dimitrova, William Lunden, and Wolfgang
  Ketterle.
\newblock Interaction spectroscopy of a two-component mott insulator.
\newblock {\em Phys. Rev. A}, 99:033612, Mar 2019.

\bibitem{FeshbachData2010}
Cheng Chin, Rudolf Grimm, Paul Julienne, and Eite Tiesinga.
\newblock Feshbach resonances in ultracold gases.
\newblock {\em Rev. Mod. Phys.}, 82:1225--1286, Apr 2010.

\bibitem{NaRb(2022)}
Zhichao Guo, Fan Jia, Bing Zhu, Lintao Li, Jeremy~M. Hutson, and Dajun Wang.
\newblock Improved characterization of feshbach resonances and interaction
  potentials between $^{23}\mathrm{Na}$ and $^{87}\mathrm{Rb}$ atoms.
\newblock {\em Phys. Rev. A}, 105:023313, Feb 2022.

\bibitem{39K87Rb(2008)}
Andrea Simoni, Matteo Zaccanti, Chiara D'Errico, Marco Fattori, Giacomo Roati,
  Massimo Inguscio, and Giovanni Modugno.
\newblock Near-threshold model for ultracold krb dimers from interisotope
  feshbach spectroscopy.
\newblock {\em Phys. Rev. A}, 77:052705, May 2008.

\bibitem{41K(2009)}
T.~Kishimoto, J.~Kobayashi, K.~Noda, K.~Aikawa, M.~Ueda, and S.~Inouye.
\newblock Direct evaporative cooling of $^{41}$K into a Bose-Einstein
  condensate.
\newblock {\em Phys. Rev. A}, 79:031602, Mar 2009.

\bibitem{PotassiumMixtures}
L.~Tanzi, C.~R. Cabrera, J.~Sanz, P.~Cheiney, M.~Tomza, and L.~Tarruell.
\newblock Feshbach resonances in potassium bose-bose mixtures.
\newblock {\em Phys. Rev. A}, 98:062712, Dec 2018.

\bibitem{cesium(1999)}
Vladan Vuleti\ifmmode~\acute{c}\else \'{c}\fi{}, Andrew~J. Kerman, Cheng Chin,
  and Steven Chu.
\newblock Observation of low-field feshbach resonances in collisions of cesium
  atoms.
\newblock {\em Phys. Rev. Lett.}, 82:1406--1409, Feb 1999.

\bibitem{85Rb133Cs}
Hung-Wen Cho, Daniel~J. McCarron, Michael~P. K\"oppinger, Daniel~L. Jenkin,
  Kirsteen~L. Butler, Paul~S. Julienne, Caroline~L. Blackley, C.~Ruth Le~Sueur,
  Jeremy~M. Hutson, and Simon~L. Cornish.
\newblock Feshbach spectroscopy of an ultracold mixture of ${}^{85}$rb and
  ${}^{133}$cs.
\newblock {\em Phys. Rev. A}, 87:010703, Jan 2013.

\bibitem{Yb(2013)}
Shinya Kato, Seiji Sugawa, Kosuke Shibata, Ryuta Yamamoto, and Yoshiro
  Takahashi.
\newblock Control of resonant interaction between electronic ground and excited
  states.
\newblock {\em Phys. Rev. Lett.}, 110:173201, Apr 2013.

\bibitem{23Na(2000)}
C.~Samuelis, E.~Tiesinga, T.~Laue, M.~Elbs, H.~Kn\"ockel, and E.~Tiemann.
\newblock Cold atomic collisions studied by molecular spectroscopy.
\newblock {\em Phys. Rev. A}, 63:012710, Dec 2000.

\bibitem{23Na(1999)}
J.~Stenger, S.~Inouye, M.~R. Andrews, H.-J. Miesner, D.~M. Stamper-Kurn, and
  W.~Ketterle.
\newblock Strongly enhanced inelastic collisions in a bose-einstein condensate
  near feshbach resonances.
\newblock {\em Phys. Rev. Lett.}, 82:2422--2425, Mar 1999.

\end{thebibliography}

\end{document}